\documentclass[prb,preprint,amsfonts,amssymb,amsmath,showpacs]{revtex4}

\usepackage{graphicx}
\usepackage{subfigure}
\usepackage{dcolumn}
\usepackage{bm}

\newcommand{\ket}[1]{\ensuremath{\left|#1\right>}}
\newcommand{\bra}[1]{\ensuremath{\left<#1\right|}}
\newcommand{\up}{\uparrow}
\newcommand{\down}{\downarrow}
\newcommand{\imag}{\mathtt{I\!m}}

\begin{document}

\title{Two-particle propagator and magnetic susceptibility in the 
Hubbard model \\ 
-An improved treatment}

\author{A. Uldry}
\email{a.uldry1@physics.ox.ac.uk}
\author{R.J. Elliott}
\email{r.elliott1@physics.ox.ac.uk}
\affiliation{Department of Physics, Theoretical Physics, 1 Keble Road, Oxford OX1 3NP, England}

\date{\today}

\begin{abstract}
We treat the two-particle Green's function in the Hubbard model using the recently developed $\tau$-CPA, a hybrid treatment that applies the coherent-potential approximation (CPA) up to a time $\tau$ related to the inverse of the band width, after which the system is averaged using the virtual-crystal approximation (VCA). This model, with suitable approximations, does predict magnetism for a modified Stoner criterion. The evaluation of the two-particle propagator in the $\tau$-CPA requires the solution of the pure CPA, within whose formalism the vertex correction and the weighted Green's functions are obtained. The dynamical susceptibility, including the vertex correction and the weighted scattering by the residual interaction, is calculated and shows a spin wave spectrum in the ferromagnetic regime.
\end{abstract}

\pacs{71.10.Fd,75.10.-b,75.40.Gb}

\maketitle

\section{\label{sec:intro}Introduction}
In the previous paper \cite{uldry} (to be referred to as I) we developed an improved treatment of the magnetic state of the simple Hubbard model, called the $\tau$-CPA. This was constructed as intermediate between the mean field Stoner model, equivalent to a VCA treatment of the energy bands, and the Hubbard's alloy analogy method, which is equivalent to the CPA. The single-particle Green's functions, 
and the properties such as density of states and magnetisation which can be developed from them, 
are given in I and will be used here without further derivations.\\
Several important magnetic properties in metallic systems, such as for example the susceptibility, depend on two-particle propagators. The treatment of these is straightforward in VCA type theories but within the CPA method a number of complications arise. In particular, because of the correlated scattering of 
the two particles from a single site, a vertex correction has to be included. The theory of such systems in the alloy situation was developed by Velick\'y \cite{velicky} in the 1960's and applied to the calculation of the electrical conduction. More recently, Elliott and Schwabe \cite{schwabe1,schwabe2} have shown how to derive weighted averages of such propagators which they applied to the theory of electron-hole pairs combined to form excitons.\\
In the present problem the susceptibility is related to the propagation of a pair of electrons of opposite spin, but because electrons of different types see different scattering centres these theories have to be generalised to the case of a four-component alloy. In addition, the residual interaction between electrons of opposite spin on the same site has to be included separately and is the dominant cause of the appearance of spin waves within the model.\\
These properties are examined in the pure Hubbard CPA case which, as has been pointed out \cite{schneidrchal,harrislange}, has only a paramagnetic phase except in special circumstances. The treatment can be carried over into the $\tau$-CPA, again requiring a matching of the propagators, for $t<\tau$ treated in the CPA with their vertex corrections, to the VCA solutions appropriate to longer times. A full treatment of this situation with further approximations to improve the tractability does indeed predict the existence of a broadened spin wave branch in the magnetic response functions.\\
The equation of motion for the two-particle propagator needed for the susceptibility 
is developed in Sec.\ref{sec:twoGFHubb}. 
Sec.\ref{sec:twoGFCPA} treats the two-particle Green's function in the pure CPA.
In Sec.\ref{sec:twoGFtau} the results are combined in order to construct the solution in the $\tau$-CPA. The dynamical susceptibility is then calculated in Sec.\ref{sec:susc}.

\section{\label{sec:twoGFHubb}Two-particle Green's functions in the Hubbard model}
As was done in I we consider the Hubbard model
\begin{equation} \label{hubbard}
H = H_0
+ U\sum_{i}c^{+}_{i\uparrow}
c_{i\uparrow}
c^{+}_{i\downarrow}c_{i\downarrow}, \quad H_{0}=\sum_{k}\epsilon_k c^{+}_{k\sigma}c_{k\sigma}
\end{equation}
for $N$ spins at temperature $T=0$.
A particular property we want to study in the Hubbard model
is the response of the system to the application of an 
external magnetic field. In the linear response theory \cite{doniach} the induced moment is given in term of the susceptibility tensor $\chi^{\mu \nu}(\mathbf{x}-\mathbf{x}',t-t')$, $\mu,\nu \in \lbrace x,y,z \rbrace$. The susceptibility tensor is completely determined by the longitudinal susceptibility 
$\chi^{zz}$ and the transverse susceptibilities $\chi^{-+}$ and $\chi^{+-}$.
We will concentrate in this paper on the transverse susceptibilities, which are given as \cite{doniach}
\begin{gather}
\chi^{-+}(\mathbf{x}-\mathbf{x}',t-t')=i 2\pi\theta (t-t')
\langle \lbrack
\sigma^{-}(\mathbf{x},t),\sigma^{+}(\mathbf{x}',t')\rbrack \rangle \label{transusc} \\
\chi^{+-}(\mathbf{x}-\mathbf{x}',t-t')=i 2 \pi\theta (t-t')
\langle \lbrack
\sigma^{+}(\mathbf{x},t),\sigma^{-}(\mathbf{x}',t')\rbrack \rangle \label{transuscPM}
\end{gather}
where $\sigma^{+}(x)=c^{+}_{x\uparrow}c_{x\downarrow}$ and 
$\sigma^{-}(x)=c^{+}_{x\downarrow}c_{x\uparrow}$.\\
The general Green's function that we want to evaluate are therefore 
of the type 
\begin{equation}\label{Kijkl}
K_{ijkl}(t) = \langle\langle c^{+}_{i\downarrow}(t)c_{j\uparrow}(t);
c^{+}_{k\uparrow}c_{l\downarrow}\rangle\rangle
\end{equation}
The transverse susceptibilities will be found as 
$\chi^{-+}(x,t)=-K_{xx00}(t)$ with a similar expression for 
$\chi^{+-}(x,t)$ given by exchanging the $\up$ and $\down$ in 
(\ref{Kijkl}).\\
The equation of motion for $K_{ijkl}$ in the Hubbard Hamiltonian 
reads
\begin{align}\label{eq_mot}
E K_{ijkl} &=
\delta_{jk}\langle c^{+}_{i\downarrow}c_{l\downarrow}\rangle
-\delta_{il}\langle c^{+}_{k\uparrow}c_{j\uparrow}\rangle  \nonumber\\
 &+  \sum_{n}t_{jn}K_{inkl}-\sum_{n}t_{ni}K_{njkl}\nonumber\\
 &+ U\langle\langle \left(
c^{+}_{i\downarrow}c_{j\uparrow}\widehat{n}_{j\downarrow}
-\widehat{n}_{i\uparrow}c^{+}_{i\downarrow}c_{j\uparrow}\right)
;c^{+}_{k\uparrow}c_{l\downarrow}\rangle\rangle_{E}
\end{align}
where $\widehat{n}_{i\sigma}$ is the usual particle number operator
$c^{+}_{i\sigma} c_{i\sigma}$. A usual procedure to close the equation of motion is to 
substitute the operators $\widehat{n}_{i\sigma}$ with numbers. In the simplest approximation the average number of spins $N^{\sigma}$ is taken. This leads to the VCA treatment 
(or Stoner model). If on the other hand the alloy analogy is applied and the operators are replaced by a random number, $0$ or $1$, we will get the CPA result for the two-particle Green's function. However, replacing the operators by numbers must be done carefully. We note that the last
term in (\ref{eq_mot}) disappears when $i=j$, as both
$c^{+}_{i\downarrow}c_{i\downarrow}c^{+}_{i\uparrow}c_{i\downarrow}$
and $c^{+}_{i\uparrow}c^{+}_{i\downarrow}c_{i\uparrow}c_{i\uparrow}$ independently go to zero. This feature is lost if only the substitution $\widehat{n}_{i\sigma}\to n_{i\sigma}$ is made, where  $n_{i\sigma}$ is either the average or $0$ or $1$ at random. The equation of motion therefore goes over to
\begin{align}\label{eq_mot_corr}
E K_{ijkl} &= 
\delta_{jk}\langle c^{+}_{i\downarrow}c_{l\downarrow}\rangle
-\delta_{il}\langle c^{+}_{k\uparrow}c_{j\uparrow}\rangle 
\nonumber\\
 &+ \sum_{n}t_{jn}K_{inkl}-\sum_{n}t_{ni}K_{njkl}\nonumber\\
 &+  U\langle\langle \left(n_{j\downarrow}-n_{i\uparrow} \right)
c^{+}_{i\downarrow}c_{j\uparrow};c^{+}_{k\uparrow}c_{l\downarrow}
\rangle\rangle_{E}\nonumber\\
 &- U\delta_{ij}\langle\langle \left(n_{j\downarrow}-n_{i\uparrow} \right)
c^{+}_{i\downarrow}c_{j\uparrow};c^{+}_{k\uparrow}c_{l\downarrow}
\rangle\rangle_{E} 
\end{align}
The term artificially introduced is thus simply subtracted again when $i=j$. \\
The evaluation of the two-particle propagator in the Hubbard model using the 
$\tau$-CPA methods require the solution of the pure CPA and VCA case. 
We will first treat the CPA case. The VCA case can be easily determined and will be discussed 
later.

\section{\label{sec:twoGFCPA}Two-particle Green's function in the pure CPA}
In this section the equation of motion (\ref{eq_mot_corr}) is treated in the alloy analogy. The numbers $n_{i\sigma}$ in (\ref{eq_mot_corr}) are now either $1$ or $0$, depending on whether there is a spin
$\sigma$ at $i$ or not. The notation $\bar{G}^{\sigma}$ will be used throughout this section instead of $G^{\sigma \, CPA}$.

\subsection{\label{subsec:fouralloy}The four-component alloy}
The alloy analogy in the two-particle case is based on the following observation: assuming that the last term in (\ref{eq_mot_corr}) can be neglected,  
the same equation of motion is obtained if the following reduced, quadratic Hamiltonian  is used instead of the Hubbard Hamiltonian:
\begin{gather}
H_{red}=\sum_{i,j,\sigma}
t_{ij}c^{+}_{i\sigma}c_{j\sigma}
+ \sum_{l,\sigma}\epsilon_{l\sigma}c^{+}_{l\sigma}c_{l\sigma} \label{redham} \\
\text{with} \quad \epsilon_{j\uparrow}= n_{j\downarrow}U \quad \text{and} \quad
\epsilon_{i\downarrow} = n_{i\uparrow}U \label{epsilons}
\end{gather}
The residual interaction, the last term of (\ref{eq_mot_corr}), will be included in the calculations at a later stage.  The rest of this section is dedicated to finding an approximation for $K^{0}_{ijkl}$, where $K^{0}_{ijkl}$ is a solution of the reduced equation of motion
\begin{align}\label{eq_mot_red}
EK^{0}_{ijkl} &= 
\delta_{jk}\langle c^{+}_{i\downarrow}c_{l\downarrow}\rangle -
\delta_{il}\langle c^{+}_{k\uparrow}c_{j\uparrow}\rangle \nonumber \\
 &+ \sum_{n}t_{jn}K^{0}_{inkl}-\sum_{n}t_{ni}K^{0}_{njkl} \nonumber \\
 &+  U \left(n_{j\downarrow}-n_{i\uparrow} \right)\langle\langle
c^{+}_{i\downarrow}c_{j\uparrow};c^{+}_{k\uparrow}c_{l\downarrow}
\rangle\rangle_{E}
\end{align}
The fact that a quadratic Hamiltonian
(\ref{redham}) is now used presents a number of advantages. For instance, it will be possible to use Wick's theorem to decompose the two-particle Green's
function into pairs of one-particle Green's functions. Moreover, it is now possible to apply the CPA formalism. 
The type of random energies
defined in the reduced Hamiltonian (\ref{redham}) suggest the
following extension of the binary alloy
analogy to the two-particle case. As seen from (\ref{epsilons}), an
up-spin travelling 
on the lattice is interacting only on the sites where there is a
down-spin. This can be interpreted in the CPA as a particle, for
example a spin up,  moving in the lattice and meeting either a
``host'' (site with no
spin down) or an ``impurity'' (site with a spin down). Since a
travelling spin up does not see the same `hosts' and `impurities'
as a spin down, we need to define the lattice as being made of four
types of sites rather than just two. Let 
$n^{+}$,$n^{-}$,$n^{\pm}$ and 
$n^{\emptyset}$ be respectively, for each site, the probability of finding a single 
electron of spin up, a single electron of spin down, both one spin up and one 
spin down, and no electron at all. The set 
$\lbrace n^{+},n^{-},n^{\pm},n^{\emptyset}\rbrace$ defines the four-component
alloy for the system. Thus $N^{\up}=n^{+}+n^{\pm}$ is the concentration of
``impurities'' seen by a down-spin and $N^{\down}=n^{-}+n^{\pm}$ 
the one seen by an up-spin. The CPA equation can be written in the form 
\cite{elliott}
\begin{equation}\label{CPAone}
\frac{N^{-\sigma}(U-\Sigma^{\sigma})}{1-(U-\Sigma^{\sigma})F^{\sigma}}
-\frac{(1-N^{-\sigma})\Sigma^{\sigma}}{1+\Sigma^{\sigma}F^{\sigma}} = 0
\end{equation}
where $F^{\sigma}$ represents the trace of $\bar{G}^{\sigma}$. 
Each equation (one for $\up$ and one for $\down$) is to be solved self-consistently for $F^{\sigma}$ and $\Sigma^{\sigma}$. The two equations
(\ref{CPAone}) are not independent, as $N^{\up}$ and $N^{\down}$ must be
 consistent together with the chemical potential for the system to be
in equilibrium.\\

\noindent The quadratic form of (\ref{redham}) allows $K^{0}_{ijkl}$ to be 
decomposed into one-particle Green's functions and correlation functions:
\begin{gather}
\langle\langle c^{+}_{i\downarrow}(t)c_{j\uparrow}(t);
c^{+}_{k\uparrow}c_{l\downarrow}\rangle\rangle = \nonumber \\
\langle c^{+}_{i\downarrow}(t)c_{l\downarrow} \rangle
\langle\langle c_{j\uparrow}(t);c^{+}_{k\uparrow} \rangle\rangle 
 +\langle c^{+}_{k\uparrow}c_{j\uparrow}(t)\rangle
\langle\langle c^{+}_{i\downarrow}(t);c_{l\downarrow}\rangle\rangle
\end{gather} 
Using the relation between time correlation functions and Green's
functions, 
$\langle
c^{+}_{i\downarrow}(t)c_{l\downarrow} \rangle$ and 
$\langle c^{+}_{k\uparrow}c_{j\uparrow}(t)\rangle$ 
 can be expressed in terms of $G^{\down}$ and $G^{\up}$, at $T=0$
\begin{equation}
\langle
c^{+}_{m\sigma}(t)c_{n\sigma}
\rangle=\int_{-\infty}^{\mu}d\omega e^{i\omega t} \left
(-\frac{1}{\pi}\mathtt{I\!m}\langle\langle c_{n\sigma};
c^{+}_{m\sigma}\rangle\rangle_{\omega} \right) 
\end{equation}
$K^{0}_{ijkl}$ can now be written in terms of the one-particle Green's
functions 
\begin{gather}
K^{0}_{ijkl}(E) =
\frac{-1}{\pi}\int_{-\infty}^{\mu}d\omega
\,G^{\up}(j,k;\omega + E)\, \imag G^{\down}(l,i;\omega) \nonumber \\
 +\frac{-1}{\pi}\int_{-\infty}^{\mu}d\omega
\,\imag G^{\up}(j,k;\omega)\, G^{\ast \down}(l,i;\omega - E)\label{Kijkl1}
\end{gather}
It is convenient to express the imaginary part of any retarded Green's
function $G^{R}(E)$  as the difference over the branch cut:
\begin{equation}
\lim_{\epsilon \rightarrow 0}\left(G\left(E+i\epsilon\right)
-G\left(E-i\epsilon\right) \right)=2\,i\,\imag G^{R}(E)
\end{equation}
(\ref{Kijkl1}) becomes 
\begin{align}\label{Kijkl2}
K^{0}_{ijkl}(E) =
\frac{i}{2\pi}&\int_{-\infty}^{\mu}d\omega
 \lbrack G^{\up}(j,k;\omega + E)\,G^{\down}(l,i;\omega)\nonumber \\
 &-G^{\up}(j,k;\omega + E)\,G^{\ast \down}(l,i;\omega) \nonumber \\
 &+G^{\up}(j,k;\omega)\, G^{\ast \down}(l,i;\omega - E)\nonumber \\
 &-G^{\ast \up}(j,k;\omega)\, G^{\ast \down}(l,i;\omega - E) \rbrack
\end{align}
This expression for $K^{0}_{ijkl}$ must now be averaged over all possible
configurations of spins. This involves the evaluation of quantities such as
$\langle G^{\up}(z1) G^{\down}(z2) \rangle$. As the particles are subject to statistical
correlations on the lattice,  $\langle G^{\up}(z1)G^{\down}(z2)
\rangle \neq\langle G^{\up}(z1)\rangle\langle G^{\down}(z2) \rangle$. 
The correction to $\langle G^{\up}(z1)\rangle\langle G^{\down}(z2) \rangle$ is expected to be small, as the up-spins do not propagate on the same lattice of impurities as the down-spins. 

\subsection{\label{subsec:vertex}Vertex corrections for the four-component alloy}
In this part the vertex correction is developed for a
four-component alloy that is consistent with the CPA. This is a straightforward
generalisation of the binary alloy case obtained by Velick\'y \cite{velicky}.The vertex corrections arise from the statistical correlations felt by the two spins as they move on the same lattice. We adopt Velick\'y's procedure and look into
evaluating $G^{(2)}(z_1,z_2):=\langle G^{\up}(z_1) C  G^{\down}(z_2) \rangle$,
where $C$ is a general operator. It will be later restricted to
diagonal operators, since for the magnetic susceptibility only quantities of the type $\langle G^{\up}_{l,m}(z_1)  G^{\down}_{m,l}(z_2) \rangle $ are 
needed. The result, after directly following Velick\'y's algebra, can be summarised as follows
\begin{equation}\label{Kkz1z2velick}
G_k^{(2)}(z_1,z_2)=\frac{a_k(z_1,z_2)}{1-\Lambda(z_1,z_2)A_k(z_1,z_2)} 
\end{equation}
where
\begin{gather}
a_k(z_1,z_2)=\sum_{m}e^{-ikR_m}\bra{m}\bar{G}^{\up}(z_1)C\bar{G}^{\down}(z_2)\ket{m} \label{dekak} \\
A_k(z_1,z_2)=\sum_{m}e^{-ikR_m}F^{\up}_m(z_1)F^{\down}_m(z_2)\label{defAk} \\
F_{n-m}:=\bra{n}\bar{G}\ket{m} 
\end{gather}
can be written in products of the single-particle functions. The vertex correction 
\begin{equation}\label{lambdadef}
\Lambda(z_1,z_2)=\frac{\langle t_n^{\up}(z_1)t_n^{\down}(z_2) 
\rangle}{1+F^{\up}(z_1)
\langle t_n^{\up}(z_1)t_n^{\down}(z_2) \rangle F^{\down}(z_2) } 
\end{equation}
gives the correlated scattering of the two particles for the same sites. 
We recall that the CPA result for the $t_n$'s is
$t_n(z)=\left( \epsilon_n-\Sigma(z) \right)[
1-\lbrack\epsilon_n-\Sigma(z)\rbrack F(z)]^{-1}$. 
If $C$ is now restricted to a diagonal operator, $A_k$ becomes equivalent to $a_k$, and they represent the free-particle solution
\begin{equation}\label{akexp}
a_k(z_1,z_2)=\frac{1}{N}\sum_q\bar{G}^{\up}_{q+k}(z_1) 
\bar{G}^{\down}_{q}(z_2)
\end{equation}
The evaluation of $\Lambda$ for the four-component alloy follows again on the same lines as for the binary alloy in the single-site representation. The average
$\langle t_n^{\up}(z_1)t_n^{\down}(z_2) \rangle$ is obtained by
summing the contributions from the four species on the lattice. To
simplify the notation a little the energies $z_1$ and $z_2$ have been
omitted again:
\begin{widetext}
\begin{align}\label{tntn}
\langle t_n^{\up}t_n^{\down} \rangle &= 
n^{+}\frac{
\left(-\Sigma^{\up}\right)\left(U-\Sigma^{\down}\right)}
{
\lbrack 1+\Sigma^{\up}F^{\up} \rbrack
\lbrack 1-\left( U-\Sigma^{\down} \right) F^{\down} \rbrack}
+
n^{-}\frac{
\left( U-\Sigma^{\up}\right)
\left(-\Sigma^{\down}\right)}
{
\lbrack 1-\left( U-\Sigma^{\up}\right)F^{\up} \rbrack
\lbrack 1+\Sigma^{\down} F^{\down} \rbrack}  \nonumber \\
 &+n^{\pm}\frac{\left
( U-\Sigma^{\up}\right)\left(U-\Sigma^{\down}\right)}
{\lbrack 1-\left( U-\Sigma^{\up}\right)F^{\up} \rbrack
\lbrack 1-\left( U-\Sigma^{\down} \right) F^{\down} \rbrack}
+n^{0}\frac{\left(-\Sigma^{\up}\right)\left(-\Sigma^{\down}\right)}
{\lbrack 1+\Sigma^{\up}F^{\up} \rbrack
\lbrack 1+\Sigma^{\down} F^{\down} \rbrack} 
\end{align}
\end{widetext}
This expression can be simplified by substituting the CPA result 
$F^{\sigma}= \left( \Sigma^{\sigma}-N^{-\sigma}U\right)
[\Sigma^{\sigma}\left(U-\Sigma^{\sigma} \right) ]^{-1}$ in (\ref{tntn}). Defining
\begin{equation}\label{ndef}
\bar{n}:=n^{\pm}-N^{+}N^{-} \quad
\eta:=\frac{\bar{n} }
{N^{+}N^{-}\left(1-N^{+}\right)\left(1-N^{-}\right) }
\end{equation}
we find
\begin{equation}\label{tntn2}
\langle t_n^{\up}t_n^{\down}
\rangle=\frac{\Sigma^{\up}\Sigma^{\down}\left(U-\Sigma^{\up} \right) 
\left(U-\Sigma^{\down}\right)}{U^2} \eta 
\end{equation}
Putting (\ref{tntn2}) into (\ref{lambdadef}) leads to the vertex correction
\begin{equation}\label{lambda}
\Lambda=\frac{\Sigma^{\up}\Sigma^{\down}(U-\Sigma^{\up})(U-\Sigma^{\down})\bar{n}}
{U^{2}N^{\up}N^{\down}n^{0}+\bar{n}(\Sigma^{\up}\Sigma^{\down}
-N^{+}U\Sigma^{\up}-N^{-}U\Sigma^{\down})}
\end{equation}
We note for later reference that the vertex correction can be put in
the form
\begin{gather}
\Lambda^{-1}=\frac{1}{n^{\pm}-N^{+}N^{-}} 
\left( 
\frac{N^{+}N^{-}n^{0}}{\Sigma^{\up}\Sigma^{\down}} 
+\frac{n^{+}N^{-}(1-N^{+})}{\Sigma^{\up}(U-\Sigma^{\down})} \right. \nonumber \\
 +\left. \frac{n^{-}N^{+}(1-N^{-})}{\Sigma^{\down}(U-\Sigma^{\up})}
+\frac{n^{\pm}(1-N^{+})(1-N^{-})}{(U-\Sigma^{\down})(U-\Sigma^{\up})}
\right) \label{lambda2}
\end{gather}
In the limit $U\rightarrow 0$, $\langle t_n^{\up}t_n^{\down} \rangle=\langle t_n^{\up}
\rangle \langle t_n^{\down} \rangle=0$, so that $\Lambda$ 
tends to zero. $n^{\pm}$ takes the same value as in the purely random
case where the two particles up and down are not interacting with
anything, that is $n^{\pm}\rightarrow N^{+}N^{-}$. 
In this respect we note that
the vertex correction, with its factor $(n^{\pm}-N^{+}N^{-})$ in the
numerator, gives a measure of the deviation from the random value.\\
In the limit $U\rightarrow \infty$, $\Sigma^{\sigma}\rightarrow
(-N^{-\sigma})[F^{\sigma}]^{-1}$, so that $\Lambda \rightarrow 
\bar{n}[n^{0} F^{\up}F^{\down}]^{-1} $. In this
limit the interaction should prevent any double occupation of the same
site, so that $n^{\pm}\rightarrow 0$. The resulting $\Lambda$ is 
proportional to both the
concentration of species $\up$ and $\down$, and inversely proportional
to the proportion of vacant sites.

\subsubsection{\label{subsubsec:effect}Effect in the large $U$ limit}
The large $U$ limit is conveniently obtained in conjunction with the limit $w \rightarrow 0$, where $w$ is half the band width. This limit is to be treated with caution when dealing with the susceptibility, since no transitions are possible if the bands have no width. It is nonetheless possible to get an idea of what will be the respective weights associated with the bands in the split-band limit. In this limit, the behaviour of the Green's function in the CPA is given by \cite{schwabe1}
\begin{equation}\label{GlimUw}
\bar{G}^{\sigma}(E)=\frac{1-N^{-\sigma}}{E}+\frac{N^{-\sigma}}{E-U}
\end{equation}
The two-particle Green's function $G^{(2)}_0(E_1,E_2)=G^{\up}(E_1)G^{\down}(E_2)$ becomes, in those limits
\begin{widetext}
\begin{equation}\label{G2limUw}
G^{(2)}_0(E_1,E_2)=\frac{N^{\up}N^{\down}}{(E_1-U)(E_2-U)}+
\frac{N^{\up}(1-N^{\down})}{E_1(E_2-U)}+\frac{(1-N^{\up})N^{\down}}{E_2(E_1-U)}+\frac{(1-N^{\up})(1-N^{\down})}{E_1 E_2}
\end{equation}
\end{widetext}
The transverse susceptibility $\chi^{-+}$ consists of transitions from the down-spin states to the up-spin states. Though (\ref{G2limUw}) suggests that four peaks will appear, the features of $\chi^{-+}$ strongly depend on the filling of the bands. In a less than half-filled band, the chemical potential will typically fall within the main sub-band. After the energy convolution on (\ref{G2limUw}), only the two last terms in (\ref{G2limUw}) will remain. \\
The inclusion of the vertex correction 
 (\ref{lambda}) has the effect of modifying the weights in the following manner:
\begin{widetext}
\begin{equation}\label{G2limUwcorr}
G^{(2)}(E_1,E_2)=\frac{n^{\pm}}{(E_1-U)(E_2-U)}+ 
\frac{(N^{\up}-n^{\pm})}{E_1(E_2-U)}+\frac{(N^{\down}-n^{\pm})}{E_2(E_1-U)}+\frac{1-N^{\up}-N^{\down}+n^{\pm}}{E_1 E_2}
\end{equation}
\end{widetext}
Since $n^{\pm}$ is always smaller than or equal to $N^{\up}N^{\down}$, the first and last terms of (\ref{G2limUwcorr}) are reduced compared to (\ref{G2limUw}). On the contrary, the two middle terms are enhanced. It is therefore to be expected that the vertex corrections will favour transitions from the main band to the impurity band and reduce transitions between the impurity bands (if they were allowed) or between the main bands. We note that the four weights in (\ref{G2limUwcorr}) are actually the respective concentrations of species.

\subsection{\label{sub:weights}Weighted Green's functions}
The concept of weighted Green's functions becomes necessary when different 
types of objects, such as the $A$ and
$B$ atoms for the binary alloy, are on the lattice. Weighting in the CPA means that we choose from which type of site the particle is  starting the
motion, as well as in some cases the type of site where the motion
ends. This results in a statistical weight given for each component. 
The properly weighted Green's function then allows us to
identify and analyse the effects of each component of the
alloy. Recently, Elliott and Schwabe \cite{schwabe1} calculated all the
necessary weights for the two-particle propagator in the binary
alloy. In the subsequent paper \cite{schwabe2}, they showed how to use
the weights to obtain a more refined treatment of the interacting
exciton in this context. The two-particle weighted Green's functions in the four component alloy will be required when the residual interaction in (\ref{eq_mot_corr}) is included in the calculations. Moreover, it will be shown that the one-particle weighted Green's functions can be used to determine $n^{\pm}$.

\subsubsection{\label{subsub:oneGF}Weighted one-particle Green's functions}

\begin{figure}
\includegraphics[width=8.6cm]{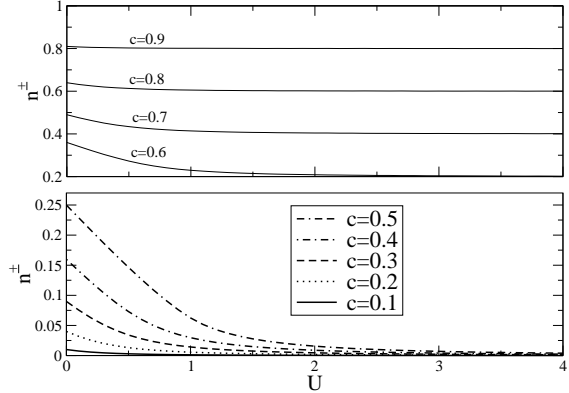}
\caption{$n^{\pm}$ for various values of the particle concentration $c$ in the CPA treatment of the Hubbard model. The state of the system is paramagnetic. \label{npmcpa}}
\end{figure}

We consider first the one-particle propagator and assume that a particle propagates between ``impurity'' sites $i$ and ``host'' sites $h$ (as it does in the four-component alloy as well). The singly-weighted Green's functions are defined such that
\begin{equation}\label{simpleweight1}
G^{io}(l,m)=
\begin{cases}
G(l,m) &\text{if l is an impurity}\\
0 & \text{otherwise}  
\end{cases}
\end{equation}
We also define $G^{oi}(l,m)$ as corresponding to the original propagator $G(l,m)$ only if $m$ is an impurity, and zero otherwise, and the doubly-weighted Green's function $G^{ii}(l,m)$ that is not zero only if both $l$ and $m$ are impurities. We define similarly $G^{ho}(l,m)$, $G^{oh}(l,m)$, $G^{ih}(l,m)$, $G^{hi}(l,m)$ and $G^{hh}(l,m)$. The weighted Green's function in an alloy of the type $H=H_0+V$ like (\ref{redham}) can be obtained simply by applying the diagonal interaction operator:
\begin{equation}\label{weightedone}
G^{io}=\frac{V}{U}G,\quad G^{oi}=G\frac{V}{U},\quad G^{ii}=\frac{1}{U^2}VGV
\end{equation}
An estimation of the averages such as $\bar{G}^{io}$ is needed in order to study how, on average, one-particle properties are made of processes on the different species of the alloy. In the CPA the one-particle single weights are 
found to be \cite{elliott,schwabe1} 
\begin{equation}\label{onesingleweight}
\bar{G^{io}}=\bar{G^{oi}}=\frac{\Sigma}{U} \bar{G} \quad
\bar{G^{ho}}=\bar{G^{oh}}=(1-\frac{\Sigma}{U}) \bar{G} 
\end{equation}
$\bar{G}^{io}$ offers a method for obtaining the average number of sites that are doubly occupied in the Hubbard model for the paramagnetic 
case. At zero temperature, the average number of sites per atom with an $\up$ spin is given by $N^{\up}=\langle c^{+}_{i\up}c_{i\up}\rangle=\int_{-\infty}^{\mu}dE
(-1/ \pi)\imag{\bar{G}^{\up}_{ii}}(E)$. 
The Green's function 
$\bar{G}^{\up}(E)$ is evaluated in the CPA. 
This formulation counts all sites where 
an $\up$ spin is present, including sites where both up and down spins are found. Recalling the alloy analogy, the $\up$ spin sees the sites where a $\down$ spin is present as defect sites $i$, where the particle interacts with 
energy $U$. The weighted Green's function $ \bar{G}^{\up\,io}_{ii}$ is thus $0$ on sites where there is no $\down$ spin, filtering out the singly occupied sites. Hence the number of sites per atoms $n^{\pm}$ where two spins are found, is, on average and in the CPA approximation
\begin{align}\label{npmOnepartCPA}
n^{\pm} &=\int_{-\infty}^{\mu}dE\left(-\frac{1}{\pi} \right)\imag{\bar{G}^{\up\,io}_{ii}}(E)\\
 &=\int_{-\infty}^{\mu}dE\left(-\frac{1}{\pi} \right)\imag{\lbrack F^{\up}(E)\frac{N^{\down}}{1-(U-\Sigma^{\up}(E))F^{\up}(E)}\rbrack } \nonumber
\end{align}    
In the limit $U\rightarrow 0$, $n^{\pm}\rightarrow N^{\up}N^{\down}$, as expected. The results for $n^{\pm}$ in the CPA treatment and for the paramagnetic case, $N^{\up}=N^{\down}$, are given in Fig.\ref{npmcpa}.
For less than half-filling, that is for $c=N^{\up}=N^{\down}\leq 0.5$, $n^{\pm}$ decreases rapidly from the random value $c^2$ to $0$. At $U=1$, $n^{\pm}$ becomes almost negligible. Beyond half-filling, saturation forces a minimum number of up- and down-spins to cohabit even at large $U$, reaching a minimum value of $2c -1$ at $U\rightarrow \infty$. 
\subsubsection{\label{subsub:twoGFsing}Singly weighted two-particle Green's functions}
Weighted two-particle Green's functions are defined
identically to the one-particle Green's function. The aim is to
estimate $\langle  G_{ij}^{\up n_1n_2}G_{ji}^{\down n_2n_1}\rangle$,
where $n_1$ and $n_2$ restrict or not the type of site $i$ and $j$ can
be. The result will be naturally linked to the vertex
correction. Elliott and Schwabe \cite{schwabe1} presented the
calculations to obtain singly weighted and doubly weighted
two-particle Green's functions for a binary alloy with clearly defined
and fixed impurities $i$ and hosts $h$. They considered the possible
weighted Green's functions generated by $\langle GCG \rangle$, where
$C$ was specialised to diagonal operators or sums of diagonal
operators. For the diagonal operator they obtained  $\langle
G^{io}CG^{oi}\rangle$ 
and $\langle
G^{ii}CG^{ii}\rangle$, from which they could deduce  $\langle
G^{ho}CG^{oh}\rangle$ and  $\langle
G^{hh}CG^{hh}\rangle$. Only the following weighted Green's functions will be required for the susceptibility: $\langle G_{ij}^{\up n^{+}}G_{ji}^{\down n^{+}}\rangle$,  $\langle G_{ij}^{\up n^{-}}G_{ji}^{\down n^{-}}\rangle$,  $\langle G_{ij}^{\up n^{+}n^{+}}G_{ji}^{\down n^{+}n^{+}}\rangle$,  $\langle G_{ij}^{\up n^{-}n^{-}}G_{ji}^{\down n^{-}n^{-}}\rangle$. For completeness, the other singly and doubly weighted Green's functions will also be calculated.\\

The calculation of $\langle G^{\up n^{\pm}o}CG^{\down on^{\pm}} \rangle$ corresponds to calculating $\langle G^{\up io}CG^{\down oi} \rangle$, as $n^{\pm}$ is acting as an impurity for both the up-spin and the down-spin. The calculation 
is essentially the same to the one made by Elliott and Schwabe \cite{schwabe1} and will thus not be repeated here. The result is 
\begin{equation}\label{G2npmoonpm}
\langle G^{\up n^{\pm}o}CG^{\down on^{\pm}} \rangle=\langle G^{\up}C
G^{\down} \rangle \frac{\Sigma^{\up}\Sigma^{\down}+\Lambda D }{U^2}
\end{equation}
with $D:=1+\Sigma^{\up}F^{\up}+\Sigma^{\down}F^{\down}$. The calculation for 
$\langle G^{\up n^{-}o}CG^{\down on^{-}} \rangle$ is carried through in 
detail in appendix \ref{app:1weight} as it differs from the Schwabe cases. The other weights are obtained in the same way. The results are
\begin{equation}\label{weightGF}
\langle G^{\up\,no}C G^{\down\,o n}\rangle=\langle G^{\up\,on}C G^{\down\,no}\rangle
 =\langle G^{\up}C G^{\down}\rangle \,\xi^{n}
\end{equation}
where $n \in \{n^{\pm},n^{+},n^{-},n^{\varnothing} \}$ and the following weights were calculated:
\begin{align}
\xi^{n^{\pm}} &:=\frac{\Lambda}{(U-\Sigma^{\up})(U-\Sigma^{\down})}
\frac{n^{\pm}(1-N^{\up})(1-N^{\down})}{n^{\pm}-N^{\up}N^{\down}} \nonumber \\
\xi^{n^{-}} &:=\frac{\Lambda}{(U-\Sigma^{\up})\Sigma^{\down}}
\frac{n^{-}N^{\up}(1-N^{\down})}{n^{\pm}-N^{\up}N^{\down}} \nonumber \\
\xi^{n^{+}} &:=\frac{\Lambda}{\Sigma^{\up}(U-\Sigma^{\down})}
\frac{n^{+}(1-N^{\up})N^{\down}}{n^{\pm}-N^{\up}N^{\down}} \nonumber \\
\xi^{n^{\varnothing}} &:=\frac{\Lambda}{\Sigma^{\up}\Sigma^{\down}}
\frac{n^{\varnothing}N^{\up}N^{\down}}{n^{\pm}-N^{\up}N^{\down}} \label{singleweights}
\end{align}
\subsubsection{\label{subsubsec:doubGF}Doubly weighted two-particle Green's functions}
Elliott and Schwabe \cite{schwabe1} obtained the following result for the doubly weighted propagator 
in the binary alloy, with $C$ a diagonal operator: 
\begin{equation}\label{G2iiii}
\langle G^{\up ii}CG^{\down ii} \rangle=\langle G^{\up}C
G^{\down} \rangle \left(\frac{\Sigma^{\up}\Sigma^{\down}+\Lambda D }{U^2}\right)^2
\end{equation}
That is, the weight of the doubly-weighted Green's function is the single weight squared. This result is not valid for general non-diagonal operators $C$, as Elliott and Schwabe showed. The result (\ref{G2iiii}) also corresponds to the four-component alloy case $\langle G^{\up n^{\pm} n^{\pm}}CG^{\down n^{\pm} n^{\pm} } \rangle$. The case $\langle 
G^{\up n^{-} n^{-}}CG^{\down n^{-} n^{-} } \rangle $ is treated in detail in appendix 
\ref{app:2weight}. The result can be generalised to the other doubly weighted Green's functions. We get:
\begin{equation}\label{doubleweights}
 \langle G^{\up\,n_1 n_2}CG^{\down\,n_2 n_1}\rangle= 
\langle G^{\up}CG^{\down}\rangle \,\xi^{n_1}\xi^{n_2}
\end{equation} 
where $n_1,n_2 \in \{n^{\pm},n^{+},n^{-},n^{\varnothing}\}$.
\subsubsection{\label{remarks}Remarks about the weights}
The relation between the weights and the vertex correction becomes transparent when looking at (\ref{lambda2}). The four terms summed in $\Lambda^{-1}$ appear to be closely related to the weights, 
with (\ref{lambda2}) rewritten as $\Lambda^{-1}=\Lambda^{-1}
\left(\xi^{n^{\varnothing}}+ \xi^{n^{+}}+ \xi^{n^{-}}+\xi^{n^{\pm}} \right)$ and the weights summing up to $1$. \\
In the small $U$ limit, $\Sigma^{\sigma} \rightarrow N^{-\sigma}U$. The vertex correction (\ref{lambda}) thus simplifies to $U^2(n^{\pm}-N^{\up}N^{\down})$.
Introducing these approximations into the weights lead to
\begin{equation}\label{WeightsmallUlim}
\xi^{n^{\pm}}\rightarrow n^{\pm} \quad \xi^{n^{-}}\rightarrow n^{-} \quad  \xi^{n^{+}}\rightarrow n^{+} \quad  \xi^{n^{\varnothing}}\rightarrow n^{\varnothing} 
\end{equation}
In the large $U$ limit the weights retain their energy dependence and estimations are of little practical use. It is however interesting to note that using the large $U$ limits
$\Sigma^{\sigma}\rightarrow
\-N^{-\sigma} [F^{\sigma}]^{-1}$ and $\Lambda \rightarrow (n^{\pm}-N^{+}N^{-})[n^{0}F^{\up}F^{\down}]^{-1}$, we get ($F^{\up}$ and $F^{\down}$ are bounded)
\begin{gather}\label{WeightbigUlim}
\xi^{n^{\pm}} \rightarrow \frac{n^{\pm}(1-N^{\up})(1-N^{\down})}{n^{\varnothing} (F^{\down}U+N^{\up})(F^{\up}U+N^{\down}) }\rightarrow 0 \quad \xi^{n^{\varnothing}} 
\rightarrow 1 \nonumber \\
\xi^{n^{\sigma}}\rightarrow -\frac{n^{\sigma}(1-N^{\sigma})}{n^{\varnothing}
(F^{-\sigma}U+N^{\sigma})} \rightarrow 0
\end{gather}
where $\sigma \in \{ +,- \}$ or $ \{ \up,\down \}$ accordingly.
In the infinitely large $U$ limit the weight is thus solely carried by the contribution from the empty sites. Contributions from occupied sites are eliminated as the repulsive interaction grows, preventing encounters of the particles on those sites when $N_e <1$.

\subsection{\label{residexp}The residual interaction}
In this subsection we turn our attention to the last term in (\ref{eq_mot_corr}). 
So far we have found the solution $K^{0}_{ijkl}$ of the reduced equation of motion. The simplest way of treating the term $-U\delta_{ij}\langle\langle \left(n_{j\downarrow}-n_{i\uparrow} \right)
c^{+}_{i\downarrow}c_{j\uparrow};c^{+}_{k\uparrow}c_{l\downarrow}
\rangle\rangle_{E}$ is to average the $n_{j\sigma}$'s to $N^{\sigma}$'s, so that it becomes $-Um\,\delta_{ij}\langle\langle c^{+}_{i\downarrow}c_{j\uparrow};c^{+}_{k\uparrow}c_{l\downarrow}
\rangle\rangle_{E}$, with $m=N^{\down}-N^{\up}$. This is the VCA treatment of this term. Calling $K_{ijkl}$ the solution of the whole equation of motion (\ref{eq_mot_corr}), it can be checked that $K_{xxx'x'}$ can be expressed in terms of the uncorrected solution by
\begin{align}
K_{xxx'x'}(&t) = K^{0}_{xxx'x'}(t) \nonumber \\
 &-U\sum_{\bar{n}}\int_{-\infty}^{\infty}dt'
K^{0}_{xx\bar{n}\bar{n}}(t-t')K_{\bar{n}\bar{n}x'x'}(t')\label{standtimeexp}
\end{align}
Fourier transforming both in time and space, we get
\begin{equation}\label{standexp}
K(q,E)=\frac{K^{0}(q,E)}{1+UK^{0}(q,E)}
\end{equation}
This result, however, is not consistent with the one-particle Green's functions obtained 
with the CPA, as (\ref{standexp}) predicts an instability in the ground state 
at $U=-1/K^{0}(0,0)$. Since the CPA does not allow for ferromagnetic solutions, a treatment of the extra term is required that is consistent with the original method.\\
A more refined treatment can be obtained when $N^{\up}\neq N^{\down}$ if the extra term is treated using the weighted Green's functions developed in the CPA. Instead of taking their averages, the operators $n_{i\sigma}$ are again replaced by $0$ or $1$ at random. The integral equation including the extra term now reads
\begin{align}
&K_{xx'}(t)= K^{0}_{xx'}(t) \nonumber \\
&-\frac{U}{m}\sum_{\bar{n}}
\left(n_{\bar{n}\down}- n_{\bar{n}\up} \right)
\int_{-\infty}^{\infty}dt'
K^{0}_{x\bar{n}}(t-t')K_{\bar{n}x'}(t') \label{weighttimeexp}
\end{align}
where to simplify the notation, we have written $K^{0}_{xx'}$ instead of 
$K^{0}_{xxx'x'}$. Transformed in the energy space, it can be developed into
\begin{gather}
K_{xx'}(E)=K^{0}_{xx'}(E)-\frac{U}{m}\sum_{\bar{n}}
K^{0}_{x\bar{n}}(E)\left(n_{\bar{n}\down}- n_{\bar{n}\up} \right)
K^{0}_{\bar{n}x'}(E) \nonumber \\
+\left(\frac{U}{m}\right)^2\sum_{\bar{n}_1,\bar{n}_2}
K^{0}_{x\bar{n}_1}(E)\left(n_{\bar{n_1}\down}- n_{\bar{n_1}\up} \right)
K^{0}_{\bar{n_1}\bar{n_2}}(E)\nonumber \\
\times \left(n_{\bar{n_2}\down}- n_{\bar{n_2}\up} \right)K^{0}_{\bar{n_2}x'}(E)
+ \dots \label{weightexp}
\end{gather}
The terms $\left(n_{\bar{n}\down}- n_{\bar{n}\up} \right)$ can take three possible values, $0$, $1$ or $-1$, depending on whether the site $n$ is an empty or doubly occupied site, a single $\down$ site, or a single $\up$ site, respectively. The summations over $\bar{n}$, $\bar{n_1}$, $\bar{n_2}$ etc in (\ref{weightexp}) can thus be restricted to singly-occupied sites. For example, the first summation in (\ref{weightexp}) becomes
\begin{gather}\label{line1weightexp}
-\frac{U}{m}\sum_{\bar{n}}
\left(n_{\bar{n}\down}- n_{\bar{n}\up} \right)
K^{0}_{x\bar{n}}(E)K^{0}_{\bar{n}x'}(E) \nonumber\\
=-\frac{U}{m}\sum_{\bar{n}\in \down}K^{0}_{x\bar{n}}(E)K^{0}_{\bar{n}x'}(E)+\frac{U}{m}\sum_{\bar{n}\in \up}K^{0}_{x\bar{n}}(E)K^{0}_{\bar{n}x'}(E)
\end{gather} 
The restricted summation of full Green's functions can be replaced by a 
summation over all sites of the weighted Green's functions. For example,
\begin{align}
-\frac{U}{m}\sum_{\bar{n}\in \down}&K^{0}_{x\bar{n}}(E)
K^{0}_{\bar{n}x'}(E) \nonumber \\
 &= -\frac{U}{m}\frac{1}{n^{-}}\sum_{\bar{n}}K^{0\,0 n^{-} }_
{x\bar{n}}(E)K^{0\,n^{-} 0}_{\bar{n}x'}(E)
\end{align}
where the summation has to be renormalised by the impurity concentration, as the averaged weighted Green's function is the total counted on all impurity sites. 
We recall that $K^{0\,0 n^{-} }_{x\bar{n}}(E)$ represents an energy convolution over $G^{0 n^{-}}_{x\bar{n}} G^{n^{-}0}_{\bar{n} x} $. 
Calling $v_{\up} \equiv U/(m \, n^{+})$ and $v_{\down} \equiv -U/(m\, n^{-})$, the expansion (\ref{weightexp}) becomes, after Fourier transforming
\begin{gather}
K(q,E)=K^{0}(q,E)+\sum_{s\in\{\up,\down \}}v_s K^{0,os}(q,E)K^{0,so}(q,E)\nonumber \\
+\sum_{s_1,s_2\in\{\up,\down \}}
v_{s_1} v_{s_2} K^{0,os_1}(q,E)K^{0,s_1s_2}(q,E)K^{0,s_2o}(q,E)\nonumber \\
+\dots \label{weightexp3}
\end{gather}
Such an expansion is very similar to what is done in the case of the excitonic absorption in a binary alloy \cite{schwabe2}. In this paper, Elliott and Schwabe calculate the optical absorption of an exciton, taking into account the interaction between the hole and the electron. This is done by writing a weighted scattering expansion comparable to (\ref{weightexp3}). Similarly to 
Ref.\onlinecite{schwabe2}, (\ref{weightexp3}) can be expressed in a compact matrix form. The following matrices are defined, where all weighted Green's functions are understood to depend on $q$ and $E$:
\begin{gather}
\vec{K^{0}}=
\begin{pmatrix}K^{0\,n^{+}} \\ K^{0\,n^{-}}\end{pmatrix}
\quad \hat{U}=
\begin{pmatrix} v_{\up}=\frac{U}{mn^{+}} & 0 \\ 0 & v_{\down}=-\frac{U}{mn^{-}}
\end{pmatrix} \nonumber \\
\hat{K^{0}}=
\begin{pmatrix}K^{0\,n^{+}n^{+}}  & K^{0\,n^{+}n^{-}}  \\ K^{0\,n^{+}n^{-}} & K^{0\,n^{-}n^{-}} \end{pmatrix}\label{matweightdef}
\end{gather}
Since the disorder average has been taken, the following simplification can be made: $K^{0\,n^{+}n^{-}}\equiv  K^{0\,n^{-}n^{+}}$, and $K^{0\,n}\equiv K^{0\,o\,n} \equiv K^{0\,n\,o}$ for any $n$.
The scattering expansion (\ref{weightexp3}) becomes 
\begin{align}
K(q,E) &= K^{0}(q,E)+\vec{K^{0}}^{T}\hat{U}\vec{K^{0}}
+\vec{K^{0}}^{T}\hat{U}\hat{K^{0}}\hat{U}\vec{K^{0}}
+\dots \nonumber \\
 &= K^{0}(q,E)+\vec{K^{0}}^{T}\hat{U}\sum_{n=0}^{\infty}\lbrack \hat{K^{0}}\hat{U}\rbrack^n \vec{K^{0}}\label{weightexp4}
\end{align}
As (\ref{weightexp4}) represents a geometric series, the equation takes the elegant form
\begin{equation}\label{weightexp5}
K(q,E)=K^{0}(q,E)+\vec{K^{0}}^{T}\hat{U}\lbrack \mathbb{I}-\hat{K^{0}}\hat{U}\rbrack^{-1}\vec{K^{0}}
\end{equation}
The result can be written in terms of the elements of the matrices defined in  (\ref{matweightdef}), so that (\ref{weightexp5}) goes over to
\begin{align}
K(q,E) &=K^{0}(q,E)+ \nonumber \\
 &\frac{v_{\down}(K^{0\,n^{-}})^2+v_{\up}(K^{0\,n^{+}})^2 +v_{\up}v_{\down} \widetilde{K}_1}{1-
v_{\down}K^{0\,n^{-}n^{-}}-v_{\up}K^{0\,n^{+}n^{+}}+
v_{\up}v_{\down}\widetilde{K}_2 } \label{cpa2weighted0}
\end{align}
with the following definitions: 
\begin{align}
\widetilde{K}_1 &=2K^{0\,n^{+}}K^{0\,n^{-}}K^{0\,n^{+}n^{-}} \nonumber \\
 &-(K^{0\,n^{-}})^2 K^{0\,n^{+}n^{+}}-(K^{0\,n^{+}})^2 K^{0\,n^{-}n^{-}} \label{tildeK1def}\\
\widetilde{K}_2 &=K^{0\,n^{+}n^{+}} K^{0\,n^{-}n^{-}}-(K^{0\,n^{+}n^{-}})^2 \label{tildeK2def}
\end{align}
This expression can be greatly simplified if the small $U$ limit is taken. In that case, the $U^2$-terms are neglected, and the weights are 
energy-independent (cf \ref{WeightsmallUlim}). Using the property (\ref{doubleweights}) for the double-weights, (\ref{cpa2weighted0}) becomes  
\begin{equation}\label{weightedUsmall}
K(q,E) \rightarrow \frac{K^{0}(q,E)}{1+\frac{U}{m} K^{0}(q,E)\lbrack \frac{(\xi^{n^{-}})^2 }{n^{-}}
- \frac{(\xi^{n^{+}})^2 }{n^{+}}\rbrack}
\end{equation}
so that the standard result (\ref{standexp}) is retrieved.\\
In the limit $U\rightarrow \infty$, we recall that the weights for $n^{+}$, 
$n^{-}$, as well as for $n^{\pm}$ are all zero. However, the weights in this case are not energy-independent and cannot be taken out of the convolution represented by $K^{0\,n^{+}n^{+}}(q,E)$ or $K^{0\,n^{-}n^{-}}(q,E)$, and the behaviour is difficult to predict.

\subsection{\label{subsec:CPApara}The paramagnetic case}
As the CPA does not allow for ferromagnetic ground
states, the paramagnetic case where $n^{+}=n^{-}$, $N^{\up}=N^{\down}$ and $m=0$ must be examined separately. In this case, the 
averaged weighted Green's functions with respect to $n^{+}$ and $n^{-}$ are expected to be identical. However, the condition that $K^{0\,on^{+}}$ is equivalent to $K^{0\,on^{-}}$ is not fulfilled by the expressions for the weights as they are given by (\ref{singleweights}). This is due to the fact that the denominator of $G^{2,n^{-}o on^{-} }(z_1,z_2)$ contains $\left( U-\Sigma^{\up}(z_1)\right)\Sigma^{\down}(z_2)$, whereas $G^{2,n^{+}o on^{+} }(z_1,z_2)$ contains $\Sigma^{\up}(z_1)\left(U- \Sigma^{\down}(z_2)\right)$. The energy convolution therefore produces two different results for  $K^{0\,n^{+}}$ and 
$K^{0\,n^{-}}$, in spite of the physical symmetry in the problem. The symmetry is however restored if $K_{-+}^{0}$ is considered together with $K_{+-}^{0}$. In the paramagnetic regime we have $K^{0\,n^{+}}_{-+}= K^{0\,n^{-}}_{+-}$, with similar equivalences for the 
doubly-weighted Green's functions. It is thus possible in principle to obtain the limit of $K_{xx}(q,E):=K_{-+}(q,E)+K_{+-}(q,E)$ when $m\rightarrow 0$.  To simplify the notation, we abandon the wave vector and energy dependences. 
The first term of the expansion of $K_{xx}$ from (\ref{weightexp3}), $K_1$,  is given by
\begin{equation}\label{term1}
K_1=\frac{U}{m}\left( \frac{ ( K_{-+}^{ n^{+}})^2} {n^{+}} -
\frac{ (  K_{-+}^{ n^{-}})^2} {n^{-}} + \frac{ (  K_{+-}^{ n^{+}})^2} {n^{+}} -
\frac{ (  K_{+-}^{ n^{-}})^2} {n^{-}} \right)
\end{equation}
The two terms in the middle and the two remaining terms on the right-hand side of (\ref{term1}) become identical in the paramagnetic limit and can be paired off. We need only consider 
one pair, for example $\lim_{m\rightarrow 0}=U/m \left( (K_{-+}^{ n^{+}})^2/n^{+} 
-(K_{+-}^{ n^{-}})^2/n^{-} \right) $. We define $\tilde{N}$ so that $N^{\up}=
\tilde{N}-m/2$ and $N^{\down}=\tilde{N}+m/2$. Similarly, $n^{+}=\tilde{n}-m/2$ and 
$n^{-}=\tilde{n}+m/2$. In the paramagnetic limit 
$ K_{-+}^{ n^{+}}=K_{+-}^{ n^{-}}
 \rightarrow K^{\tilde{n}}$.   Both $ K_{-+}^{ n^{+}}$ and $ K_{+-}^{ n^{-}}$ 
can be developed in powers of $m$ so that
\begin{equation}
\frac{(K^{n^{+}})^2}{n^{+}}=\frac{ (K^{\tilde{n}})^2}{\tilde{n}}+m\frac{\partial}{\partial \tilde{n}}
\left(\frac{(K^{\tilde{n}})^2}{\tilde{n}} \right)\frac{\partial \tilde{n}}{\partial m}
+ \mathcal{O}(m^2)
\end{equation}
The two terms together give
\begin{equation}
\frac{(K^{n^{+}})^2 }{n^{+}}-\frac{(K^{n^{-}})^2 }{n^{-}}=
-m\left(-\frac{ (K^{\tilde{n}})^2}{\tilde{n}^2} +\frac{2}{\tilde{n}}K^{\tilde{n}}\frac{\partial K^{\tilde{n}}}{\partial \tilde{n}} \right)
\end{equation} 
Due to the self-consistent nature of the CPA, the quantity 
$\partial K^{\tilde{n}}/ \partial \tilde{n} $ is very difficult to obtain. 
The self-energy varies in a complex way with the concentration of impurities, as the background itself is affected by a small modification in the number of impurities. The weights themselves are also dependent on the self-energy and the concentration of species in a complex manner. In two cases, however, the quantity $\partial K^{\tilde{n}}/ \partial \tilde{n} $ can be established. In the limit $U\rightarrow 0$ it is easy to see that 
$\partial K^{\tilde{n}} / \partial \tilde{n} \rightarrow K^{\tilde{n}} / \tilde{n}$. 
The same result is obtained in the split-band limit, since in this case 
$K^{\tilde{n}}(z_1,z_2)$ is given simply by 
$\tilde{n}/[z_1(z_2-U)]$. In the limit of small and large $U$, the first term becomes
\begin{equation}\label{solterm1}
K_1\rightarrow -\frac{U}{n^{+} n^{-}}\left(K_{-+}^{ n^{+}} \right)^2 -\frac{U}{n^{+} n^{-}}\left(K_{-+}^{ n^{-}} \right)^2 
\end{equation}
It is easy to see that the subsequent terms in $U$ can all be paired 
in the same way. 
Extrapolating the result (\ref{solterm1}) to the higher order terms, we find
\begin{equation}\label{paramapprox}
K_{xx}(q,E)\rightarrow 2 K^{0}(q,E)
-\frac{U\frac{ (K^{0\,n^{+} })^2  }{n^{+}n^{-} }} {1+U\frac{K^{0\,
n^{+}n^{+} }  }{n^{+}n^{-} } }-\frac{U\frac{ (K^{0\,n^{-} })^2}
{n^{+}n^{-}  }} {1+U\frac{K^{0\,
n^{-}n^{-} }  }{n^{+}n^{-} } }
\end{equation}
Numerical results are given in the next subsection.

\subsection{\label{numericalCPA}Numerical illustrations}
The numerical applications of the formalism developed for the CPA susceptibility are presented here. We use a three-dimensional, simple cubic band at $T=0$. We first evaluate the susceptibility $\chi_{0}^{xx}(q,E)$ without taking into account 
the residual interaction in order to illustrate the effects of the vertex correction. The results $\chi^{xx}(q,E)$ that include the residual interaction are discussed next.\\

\begin{figure}
\includegraphics[width=8.6cm]{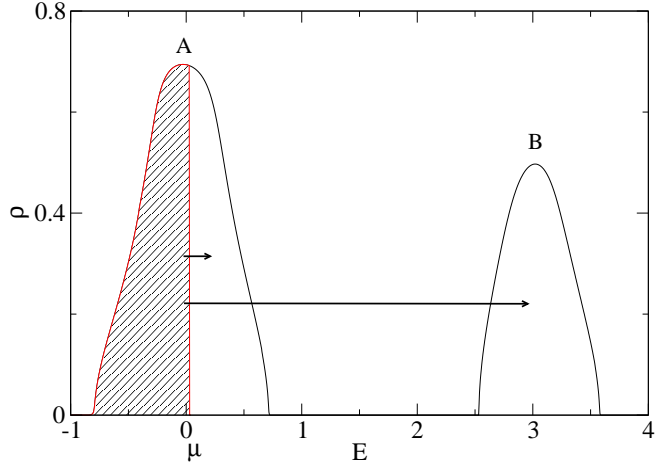}
\caption{DOS (CPA) for U=3.0, c=0.35 and $\mu=0.02$.}\label{figdoscpaU3c35}
\end{figure}

\begin{figure}
\subfigure[$U=0.5$, $n^{\pm}=0.04$ and $\mu=-0.05$ ]{\label{cpapara1}
\includegraphics[width=8cm]{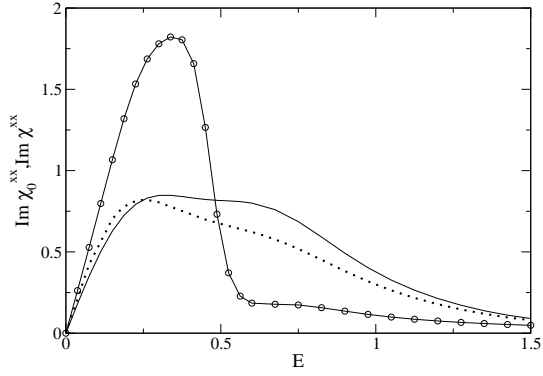}}
\subfigure[$U=1.0$, $n^{\pm}=0.02$ and $\mu=0.0$]{\label{cpapara2}
\includegraphics[width=8cm]{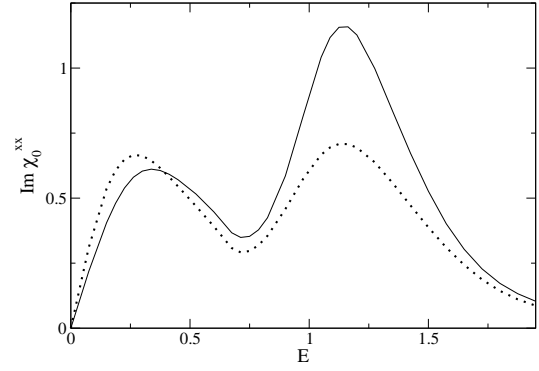}}
\subfigure[$U=3.0$, $n^{\pm}=0.0$ and $\mu=0.02$]{\label{cpapara3}
\includegraphics[width=8cm]{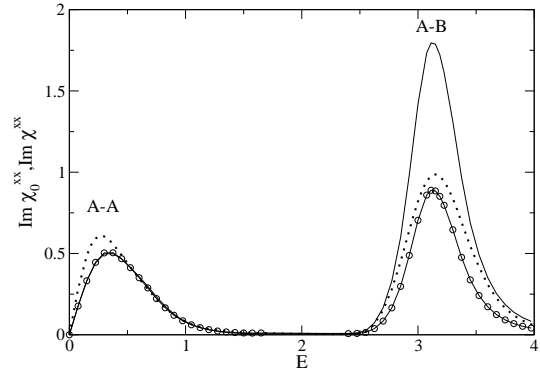}}
\caption{Imaginary part of $\chi_{0}^{xx}(q,E)$ and $\chi^{0\,xx}$ 
at $q=0$ and 
for $N_e=0.7$. Dotted line: $\chi_{0}^{xx}$ without the inclusion of the vertex correction, solid line:  $\chi_{0}^{xx}$ with the inclusion of the vertex correction, circles (when appropriate): $\chi^{xx}$ including both the vertex correction and the residual interaction.\label{figcpapara}}
\end{figure}

The susceptibility is $\chi_{0}^{xx}(q,E)=\chi_{0}^{-+}(q,E)+\chi_{0}^{+-}(q,E)$ which, in the paramagnetic ground state of the CPA, becomes simply $\chi_{0}^{xx}(q,E)=2\chi_{0}^{-+}(q,E)$. $\chi_{0}^{-+}(q,E)$ is obtained as $\chi_{0}^{-+}(q,E)=-K^{0}(q,E)$, where $K^{0}(q,E)$ is calculated from the combination of Green's functions (\ref{Kijkl2}). Each of these Green's functions is obtained from the CPA solution (\ref{Kkz1z2velick}) with the vertex correction (\ref{lambda}). The quantity $n^{\pm}$ is determined from the weighted Green's functions (\ref{npmOnepartCPA}). The transitions involved in calculating the Green's function $K^{0}(q,E)$, at $q=0$ and without the inclusion of the vertex corrections, are illustrated in Fig.\ref{figdoscpaU3c35}. It shows the density of states in a split-band case, $U=3.0$, 
$N^{+}=N^{-}=0.35$ and $\mu=0.02$.
In this paramagnetic case the up-spin band is identical to the down-spin band. Transitions occur between the occupied states (shaded area in Fig.\ref{figdoscpaU3c35}) of the down-spins and the empty states of the up-spins. As the chemical potential falls within the lower sub-band A, transitions start at zero energy $E$. A first peak in the imaginary part of $\chi_{0}^{xx}(q=0,E)$ thus extends from $E=0$ 
to the size of the sub-band A, approximatively $E=1.5$ is this case. Transitions can also occur at higher energies between the band A and the impurity 
band B. A second peak is thus created that starts at an energy equivalent to the distance between band B and the chemical potential $\mu$ and ends at an energy equivalent to the distance between the higher edge of B and the lower edge of A. The result in this case is shown as the dotted line in 
Fig.\ref{cpapara3}. $U$ is large enough in this case for the large 
$U$ limit (\ref{G2limUw}) to be applicable. From the last two terms in (\ref{G2limUw}) we can estimate the weight of the A-A transitions compared to the weight of the A-B transactions. Since the main sub-band is only half-filled, the ratio A-A/A-B is approximatively $0.9$.
Examples for smaller $U$ are given in Fig.\ref{cpapara1} 
and \ref{cpapara2} (dotted lines).
At $U=1.0$ and $U=0.5$ ($N_e=0.7$), the two peaks can still be identified even though they are not separated by a gap. \\

The effects of the vertex correction are depicted in 
Fig.\ref{figcpapara} (solid line) for small, medium and large $U$, 
compared to the non corrected $\imag{\chi_{0}^{xx}}(q=0,E)$ (dotted line). As predicted by (\ref{G2limUwcorr}), the A-A transitions are slightly reduced while the A-B transitions are noticeably enhanced. This is to be understood from the perspective of the spin up and down propagating on the four-component alloy. As the spin up interacts with the down-spins and the spin down with the up-spins, it is to be expected that the $n^{\pm}$ sites will play a major role in correlating the particles motions. For both the propagating spin up and down, the $n^{\pm}$ sites are included in the impurity B sub-band, hence enhancing transitions of the A-B type. 
Similarly, the $n^{o}$ sites, with which no spin interacts, 
are included in the A sub-band of each species. 
Compared to the free particle cases, it is thus expected that the inclusions of correlations will reduce the transitions from one A sub-band to the other. \\

The result of the approximation (\ref{paramapprox}) for $\chi^{xx}(E)$ at $q=0$ that includes the residual interaction is plotted (circles) in its range of validity: Fig.\ref{cpapara1} for $U=0.5$, $N_e=0.7$ and $n^{\pm}=0.04$, and for a split-band case at $U=3$, $N_e=0.7$ and $n^{\pm}=0.0$, Fig.\ref{cpapara3}. 
Whereas the inclusion of the vertex corrections generally reduces the A-A peak and enhances the A-B peak, the correction (\ref{paramapprox}) has different effect at small and large $U$. At small $U$ the A-A transitions are enhanced compared to the value including only the vertex correction. At large $U$ the A-A transitions have come back to their vertex correction value. In both cases though, the A-B transitions are strongly reduced.
The approximation (\ref{paramapprox}) is not valid at intermediate $U$. It in fact causes the imaginary part of $\chi^{xx}(E)$ to change sign between the two bands, hence 
producing unphysical results in this range. Therefore no results for $\chi^{xx}(E)$ were plotted in Fig.\ref{cpapara2}\\

\begin{figure}
\includegraphics[width=8.6cm,height=!]{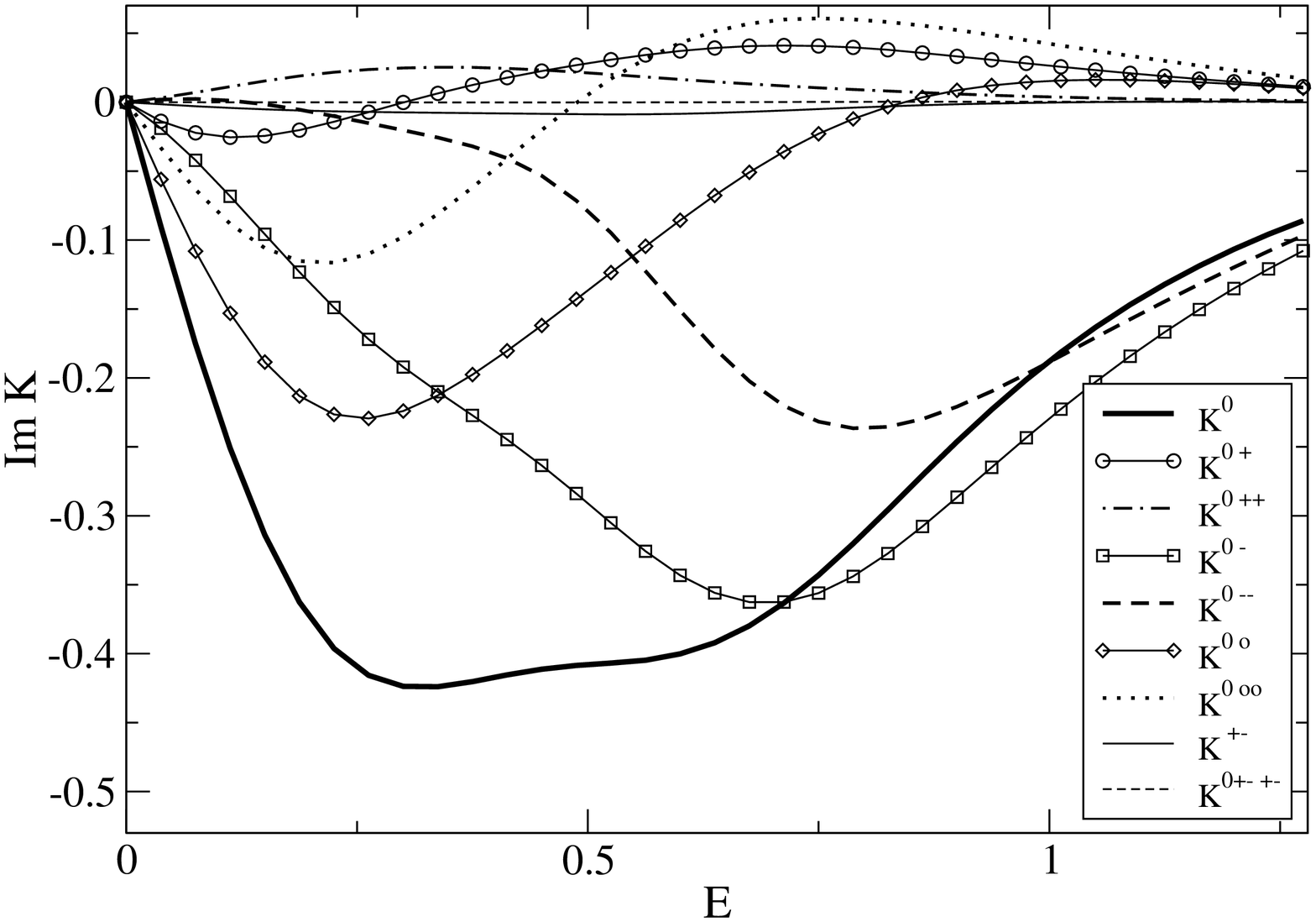}
\caption{Imaginary part of the singly- and doubly- weighted Green's functions compared to the non-weighted case for $U=0.5,N^{\up}=N^{\down}=0.35, n^{\pm}=0.04,\mu=-0.05$.}\label{figweights05}
\end{figure}

\begin{figure}
\includegraphics[width=8.6cm,height=!]{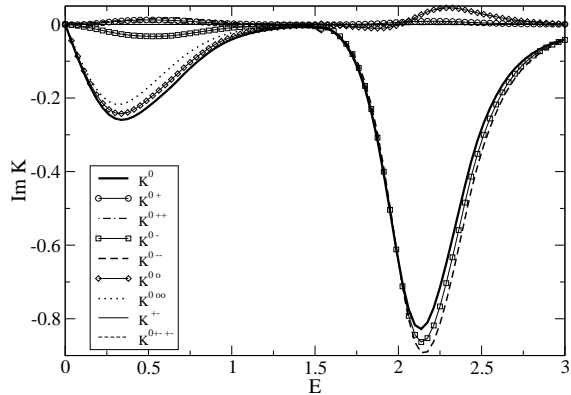}
\caption{Imaginary part of the singly- and doubly- weighted Green's functions compared to the non-weighted case for $U=2.0,N^{\up}=N^{\down}=0.35, n^{\pm}=0.0, \mu=0.02$.}\label{figweights20}
\end{figure}

An example of the singly and doubly weighted Green's functions obtained for the susceptibility is given in Figs.\ref{figweights05} and \ref{figweights20}, for $U=0.5$ and $U=2.0$. In these pictures, $K_{-+}^{0}:=K^{0}$, $K^{0\,+}:=K^{0\,n^{+}}$ etc. We recall that $\chi^{0\,xx}=-2 K_{-+}^{0}$ in the paramagnetic case. By construction, $K^{0}=K^{0+}+K^{0-}+K^{0o}+K^{0\pm}$. Whereas $\chi^{0\,xx}$ is an observable quantity and is therefore positive definite (and accordingly, the imaginary part of $K^{0}$ is always negative), nothing can be said about the weighted quantities. As can be seen in Figs.\ref{figweights05} and \ref{figweights20}, some weighted Green's functions contribute negatively for a part or for the whole spectrum. The $n^{\emptyset}$ Green's functions contribute negatively to the A-B peak at all strengths of $U$. At low $U$, $K^{0+}$ is negative in the A-A band, and positive in the A-B band. At large $U$, the contribution from the $n^{+}$ quantities becomes comparatively small. For both the small $U$ and large $U$ case, the A-A peak is made up mostly of $K^{0\,o}$ and $K^{0\,oo}$, that is, from the $n^{\emptyset}$ contributions. Similarly, the A-B peak is constituted mainly of the $K^{0 -}$ contributions. $K^{0 \pm}$ and $K^{0 \pm \pm }$, due to the size of $n^{\pm}$, 
do not contribute significantly. \\

Whereas the picture given by the singly weighted Green's functions for $\chi_{-+}$ and $\chi_{+-}$ corresponds well to the physical reality, the weighted residual interaction in the paramagnetic case can only be considered for $\chi_{xx}$. The limit $m \rightarrow 0$ was not obtained in a general case. Instabilities in the ground state, if they can occur, would happen when $K^{dd}(q=0,E=0)
=-n^{+}n^{-}/U$, with $d\in \{+,-\}$. $K^{dd}(q=0,E=0)$ is going to zero when $U$ increases. In the large $U$ limit, we find
\begin{equation} 
G^{(2) + +}(\omega,\omega)\rightarrow \frac{(n^{+})^2}{\omega^2 -U \omega (2-N^{\up}-N^{\down}) +n^{\emptyset}U^2}
\end{equation}
This leads us to think that $K^{dd}(q=0,E=0)$ goes to zero with $1/U^2$, so that in agreement with the one-particle result, no instabilities should be found in the ground state.\\

The approximation (\ref{paramapprox}) is only valid in both extreme $U$ limits and cannot be interpolated at intermediate $U$. However, it can be predicted that the peak at the energy of the maximum of the A-B transition will be reduced from its value obtained from the vertex correction. This is an expected behaviour considering the way we split the equation of motion. In the first part of the problem, when considering the propagation of two particles on a four-component alloy, the inclusion of the vertex correction led to the enhancement of the interband transitions. Due to the nature of the problem, it is indeed expected for the particles to become correlated when they meet on an $n^{\pm}$ site, the type of site that represents an impurity for both of them. However, the Pauli principle forbids the $\up$ and $\down$ spin to flip on such a site. When considering one-particle Green's functions, the Pauli principle is taken into account through the Fermi statistics. The vertex correction is obtained from the point of view of two particles moving in a lattice of potentials, rather than an up and down spins moving among other up and down spins. It is thus not clear whether the inclusion of the vertex correction conserves the Pauli principle or not. The inclusion of the last term in the equation of motion in the form of a weighted scattering correction appears to adjust the results of the vertex correction so that they comply to the Pauli principle.   
 
\section{\label{sec:twoGFtau}Two-particle Green's functions in the $\tau$-CPA}
The $\tau$-CPA treatment for the one-particle Green's functions is extended to the two-particle case, by using the two-particle CPA propagator up to a time $\tau$, thereafter the VCA propagator. The resulting Green's function is thus made of a CPA part that 
includes the vertex correction and of a VCA part, both weighted by factors depending on $\tau$.\\

The two-particle Green's functions in the $\tau$-CPA can be built in a straight-forward manner from the one-particle case. The consistent way to build the two-particle Green's function $\langle G^{(2)}(t)\rangle $ is to break it in a VCA part and CPA part, as for the one-particle case:
\begin{equation}\label{twotaucpatime}
\langle G^{(2)}(t)\rangle =
\begin{cases}
\langle G^{(2)CPA}(t)\rangle  &\text{if }t\leq \tau\\
\langle G^{(2)VCA}(t)\rangle & \text{if } t> \tau  
\end{cases}
\end{equation}
For any time $t$, $\langle G^{(2(}(t)\rangle=\langle G^{\up}(t)G^{\down}(t)\rangle$. In the one-particle case, the expression for the Green's function in time could be easily obtained in the VCA case. In the CPA case, a reasonable form in time could only be assumed. As is discussed in I, this is due to the nature of the CPA poles.
To obtain the two-particle function of the type 
$\langle G^{(2)}(E)\rangle =\int_{-\infty}^{\infty}d\omega \langle G^{(2)}(\omega,E-\omega) \rangle $, where 
$G^{(2)}(\omega,E-\omega)=G^{\up}(\omega) G^{\down}(E-\omega)$, one has to evaluate
\begin{gather}
\langle G^{(2)}(E) \rangle =\int_{-\infty}^{\infty}dt\,e^{iEt}
\langle G^{(2)}(t)\rangle \\
 =\underbrace{\int_{-\infty}^{\tau}dt\,e^{iEt} \langle G^{(2)CPA}(t)\rangle }_{=:
 \langle \widetilde{G}^{(2)CPA}(E)\rangle  }+\underbrace{\int_{\tau}^{\infty}dt\,e^{iEt} \langle G^{(2)VCA}(t)\rangle }_{=: \langle \widetilde{G}^{(2)VCA}(E)\rangle } \nonumber
\end{gather}
The VCA part of the two-particle Green's function is easily obtained. 
Using the definition from I
\begin{equation}\label{VCAnat} 
\mathcal{G}_{k}^{\sigma VCA}(E)=\frac{1}{2\pi}\int_{\tau}^{\infty}dt 
\,e^{iEt}G_{k}^{\sigma VCA}(t)=
\frac{e^{i\left(E-\epsilon_k-N^{-\sigma}U  \right) \tau}}{E-\epsilon_k-N^{-\sigma}U } 
\end{equation}
the VCA part can be expressed as the convolution of the one-particle functions:
\begin{equation}\label{twoVCAnat}
\langle \mathcal{G}_{k}^{(2)VCA}(E)\rangle = \int_{-\infty}^{\infty}
d\epsilon 
\frac{1}{N}\sum_{q}\mathcal{G}_{k+q}^{\up VCA}(\epsilon) \mathcal{G}_{q}^{\down VCA}(E-\epsilon)
\end{equation}
and $\mathcal{G}^{(2)VCA}(t) =\mathcal{G}^{\up(VCA)}(t) 
\mathcal{G}^{\down VCA}(t)$. The factor for matching the two 
Green's functions at time $\tau$ is included as usual in the VCA part, so that we have
\begin{equation}\label{twoVCA}
\langle \widetilde{G}_{k}^{(2)VCA}(\epsilon_1,\epsilon_2)\rangle = 
\frac{1}{N}\sum_{q}\widetilde{G}_{k+q}^{\up VCA}(\epsilon_1) 
\widetilde{G}_{q}^{\down VCA}(\epsilon_2)
\end{equation}
with, from I, the one-particle VCA part that includes the normalisation $\beta_k$
\begin{equation}   \label{vcapart}   
\widetilde{G}^{\sigma \,VCA}_{k}(E):=\frac{1}{\beta_k}
\left(  \frac{e^{\imag{\Sigma^{\sigma}}(E)\tau}}{E -\epsilon_k-N^{-\sigma} U}
\right )
\end{equation}
The same approximation will be assumed to hold also for the CPA part. In 
that case, the procedure must be adapted to treat the vertex corrections correctly. The $\tau$-CPA says that the interaction is treated in the CPA up to a time $\tau$, from when it is treated in the VCA. In both cases, the interaction is taken in full. Therefore the vertex correction term must involve the full 
CPA Green's functions and not only the CPA equivalent of (\ref{vcapart}),
\begin{equation}\label{cpapart}
\widetilde{G}^{\sigma \,CPA}_{k}(E):=\frac{1}{\beta_k}
\left( \frac{1-e^{\imag{\Sigma^{\sigma}}(E)\tau}}
{E -\epsilon_k-\Sigma^{\sigma}(E)}\right ) 
\end{equation}
The pure CPA two-particle Green's function is given by (\ref{Kkz1z2velick}).
A similar treatment to that for the one-particle Green's functions, where the energy dependence of $\Sigma$ was ignored temporarily while the time integration was carried over, is applied here 
as well. The vertex correction to the free particle case, the denominator of 
(\ref{Kkz1z2velick}), is kept unchanged in the pure CPA form while the numerator becomes, as for the VCA case, a convolution on the $\tau$-CPA one-particle Green's functions. The following approximation will thus be used for the CPA part of the two-particle Green's function in the $\tau$-CPA:
\begin{align}
\langle \widetilde{G}_{k}&^{(2)CPA}(\epsilon_1,\epsilon_2)
\rangle = \nonumber \\
 &\frac{\frac{1}{N}\sum_{q}\widetilde{G}_{k+q}^{\up CPA}(\epsilon_1)
\widetilde{G_{q}}^{\down CPA}(\epsilon_2)}
{1-\Lambda(\epsilon_1,\epsilon_2)\frac{1}{N}\sum_{q} G_{k+q}^{\up CPA}
(\epsilon_1)G_{q}^{\down CPA}(\epsilon_2))} \label{cpataucorr}
\end{align}
$\Lambda(\epsilon_1,\epsilon_2)$ is calculated using the 
self-energy obtained from the full CPA equation.

\section{\label{sec:susc}Dynamical susceptibility in the$\tau$-CPA }
In this section we put together the elements developed in the prior sections. The two-particle Green's function calculated so far does not take into account the extra $U$-term of the equation of motion included in subsection \ref{residexp}. In order to use the formalism developed above to calculate the susceptibility, the residual interaction between the two spins must 
be included in both the VCA and CPA two-particle 
parts.

\subsection{\label{taususc}Construction}
The transverse susceptibility, without the residual term, is given by the two-particle Green's function $K_{k}^{0}(E) $, built from the disorder average of (\ref{Kijkl2}). The approximation developed so far gives
\begin{equation}\label{noexpK0tau}
K_{k}^{0\tau CPA}(E)=\widetilde{K}_{k}^{0 VCA }(E)+\widetilde{K}_{k}^{0 CPA}(E)
\end{equation}
where $\widetilde{K}_{k}^{0 VCA}(E)$ and $\widetilde{K}_{k}^{0 CPA}(E)$ are given by the convolutions (\ref{Kijkl2}) of the two-particle Green's functions (\ref{twoVCA}) and 
(\ref{cpataucorr}). \\
The inclusion of the residual $U$-term must be considered now.
The correction for a pure VCA system is given by (\ref{standexp}) whereas the pure CPA requires the weighted result (\ref{cpa2weighted0}). Once again, a straight-forward treatment would be to Fourier transform (\ref{standexp}) and (\ref{cpa2weighted0}), then to integrate the CPA part from $0$ to $\tau$ and the 
VCA part from $\tau$ to $\infty$. This would affect both the numerator and the denominator of (\ref{standexp}) and (\ref{cpa2weighted0}). The denominators contain the correction due to the interaction. It was argued in the context of the vertex correction for $\langle \widetilde{G}^{(2)CPA}(E)\rangle$ 
that the correction is kept unchanged, since the interaction must be included at all time. An identical procedure is applied here, so that the time integration is carried only on the numerators:
\begin{equation}\label{approxKkall}
K_{k}^{\tau CPA}(E)\approx \frac{\widetilde{K}_{k}^{0 VCA}(E)}{1+U K_{k}^{0 VCA}(E)} + \frac{\widetilde{K}_{2\,k}^{0 CPA}(E)}{1+K_{1\,k}^{0 CPA}(E)}
\end{equation}
$K_{k}^{0 VCA}(E)$ is  given by the convolutions (\ref{Kijkl2}) of the complete two-particle Green's functions $G_{k}^{(2)VCA}(\epsilon_1,\epsilon_2)=
N^{-1}\sum_{q}G_{k+q}^{\up VCA}(\epsilon_1)G_{q}^{\down VCA}
(\epsilon_2)$. $K_{1\,k}^{0 CPA}(E)$ and $\widetilde{K}_{2\,k}^{0 CPA}(E)$ are defined below. In order to simplify the notation, the index CPA will be abandoned for the weighted Green's function, and $n^{+}$ becomes just $+$ (with identical definitions for the other species of the alloy). All functions are evaluated at the energy $E$.
\begin{align}
K_{1\,k}^{0 CPA} &=
\frac{U}{m}\left( 
\frac{ K_{k}^{0- - }}{n^-}- 
\frac{K_{k}^{0++} }{n^{+}} \right) \nonumber\\
 &-\frac{U^2}{m^2 n^{+}n^{-}}
\left( K_{k}^{0++ } K_{k}^{0- - }
-\left(K_{k}^{0+- } \right)^2   \right) \label{defKden}
\end{align}
\begin{align}
\widetilde{K}_{1\,k}^{0 CPA} &=
\frac{U}{m}\left( 
\frac{ \widetilde{K}_{k}^{0- - }}{n^-}- 
\frac{\widetilde{K}_{k}^{0++} }{n^{+}} \right) \nonumber \\
 &-\frac{U^2}{m^2 n^{+}n^{-}}
\left( \widetilde{K}_{k}^{0++ } \widetilde{K}_{k}^{0- - }
-\left(\widetilde{K}_{k}^{0+- } \right)^2   \right)\label{defKdentilde}
\end{align}
\begin{multline}
\widetilde{K}_{2\,k}^{0 CPA}=
\widetilde{K}_{k}^{0 CPA}\left[1+
\widetilde{K}_{1\,k}^{0 CPA} \right] \\
-\frac{U}{m}\left[ \frac{ \left(\widetilde{K}_{k}^{0-} \right)^2 }{n^{-}} -\frac{ \left(\widetilde{K}_{k}^{0+} \right)^2 }{n^{+}} \right] 
+\frac{U^2}{m^2 n^{+}n^{-}} \\
\times \left[ \left( \widetilde{K}_{k}^{0+}\right)^2   
\widetilde{K}_{k}^{0--} 
+ \left( \widetilde{K}_{k}^{0-}\right)^2  
\widetilde{K}_{k}^{0++} -2 \widetilde{K}_{k}^{0+}\widetilde{K}_{k}^{0-}\widetilde{K}_{k}^{0+-} \right] \label{defKnum}
\end{multline}
This formulation is not completely consistent with the rest of the picture. The denominators of both VCA and CPA parts do not yet contain the full interaction. We recall that (\ref{Kijkl2}) is an integral from $0$ to the chemical potential $\mu$. $\mu$ is obtained consistently from the one-particle $\tau$-CPA Green's function $G^{\tau CPA}$ (see I) and not from the pure CPA or the pure VCA. The problem can be retraced back to the development (\ref{weighttimeexp}) relating $K$ to $K^{0}$. We note that the expression (\ref{weighttimeexp}) is also valid in the case of the pure VCA, since in this case the factor $\left(n_{\bar{n}\down}-n_{\bar{n}\up} \right)$ is simply averaged to $m$, consistently with the VCA. Going back to the expansion (\ref{weighttimeexp}), it is clear that the term $m$ in $U/m\sum_{\bar{n}}\left(n_{\bar{n}\down}-  n_{\bar{n}\up}\right)$ comes from the inhomogeneous term $ 2\pi \delta (t) \left( 
\delta_{jk} \langle c^{+}_{i\downarrow}c_{l\downarrow}\rangle
-\delta_{il}\langle c^{+}_{k\uparrow}c_{j\uparrow}\rangle \right) $ in the time version of the equation of motion (\ref{eq_mot}). The two correlation functions involved are thus to be calculated at the time $t=0$. If a pure VCA (CPA) system is considered, the $m$ generated by the inhomogeneous term will be called $m^{VCA}$ ($m^{CPA}$):
\begin{gather}
m^{VCA}=N^{\down VCA}-N^{\up VCA}= \nonumber\\
\int_{-\infty}^{\mu}d\epsilon \left(-\frac{1}{\pi} \right)\frac{1}{N}\sum_{q} \imag{} \lbrack G_{q}^{\up VCA}(\epsilon)
-G_{q}^{\down VCA}(\epsilon)\rbrack \label{mvcadef}\\
m^{CPA}=N^{\down CPA}-N^{\up CPA}= \nonumber \\
\int_{-\infty}^{\mu}d\epsilon \left(-\frac{1}{\pi} \right)\frac{1}{N}\sum_{q} \imag{}\lbrack G_{q}^{\up CPA}(\epsilon)
-G_{q}^{\down CPA}(\epsilon)\rbrack \label{mcpadef}
\end{gather}
$G_{k}^{\sigma VCA}$ is given by the full VCA one-particle result and 
by the full CPA (as detailed in I).
Hence, the expansion (\ref{standexp}) for the VCA case should be written as 
\begin{equation}\label{VCArecast}
K^{VCA}(q,E)=\frac{K^{0 VCA}(q,E)}{1+U\frac{m}{m^{VCA}}K^{0 VCA}(q,E)}
\end{equation}
In a pure VCA case $m$ is simply equivalent to $m^{VCA}$. 
The CPA case (\ref{cpa2weighted0}) must also be re-casted in the same way, by subsisting $m\rightarrow m^{CPA}$ in (\ref{defKden}-\ref{defKnum}).
If the case of a pure CPA is considered, $m \equiv m^{CPA}$. The approximation for the two-particle Green's function, involving the CPA vertex corrections and the extra $U$-term of the equation of motion, is thus
\begin{gather}
K_{k}^{\tau CPA}(E) = \nonumber \\
\frac{\widetilde{K}_{k}^{0 VCA}(E)}{1+U\frac{m}{m^{VCA}}K_{k}^{0 VCA}(E)}+
\frac{\widetilde{K}_{2\,k}^{0 CPA}[m^{CPA}](E)}
{1+ K_{1\,k}^{0 CPA}[m^{CPA}](E)}\label{Kkall}
\end{gather}
The chemical potential $\mu$, being obtained from the whole $G^{\tau CPA}$, 
will in general lead to an $m^{CPA}$ and $m^{VCA}$ quite different from the 
values obtained in pure systems. \\ 

It is crucial to keep the full interaction, as it guarantees the appearance of the spin waves at $E=0$ and $k=0$. As ferromagnetism sets in, collective 
spin wave excitations are expected to appear. The magnetisation at $k=0$ can assume any direction in the averaged system and the energy cost $E$ for exciting spin waves must be zero. This is also found in the $\tau$-CPA. The spin wave spectrum is generated almost entirely from the VCA part. As will be illustrated in the following paragraph, the CPA part at large $U$ is negligible at low energy $E$. But even at lower $U$ we recall that the CPA part does not create instabilities at low energy, so only the VCA part of (\ref{Kkall}) is considered for the moment. 
In a ferromagnetic ground state, the following limit is obtained:
\begin{equation}\label{chi0limit}
 \lim_{k\rightarrow 0} K^{0 VCA}_{k}(E) \rightarrow \frac{m^{VCA}}{E-Um} 
\end{equation} 
Putting (\ref{chi0limit}) into (\ref{VCArecast}), the result for finite $E$ and $k\rightarrow 0$ is
\begin{equation}\label{chilimit}
 \lim_{k\rightarrow 0} K^{VCA}_{k}(E) \rightarrow \frac{m^{VCA}}{E} 
\end{equation}
For small but finite $k$, we expect the excitation energy $E$ to increase. As in the Stoner model, the spin waves will finally merge with the continuum. \\

\begin{figure}
\includegraphics[width=8.6cm,height=!]{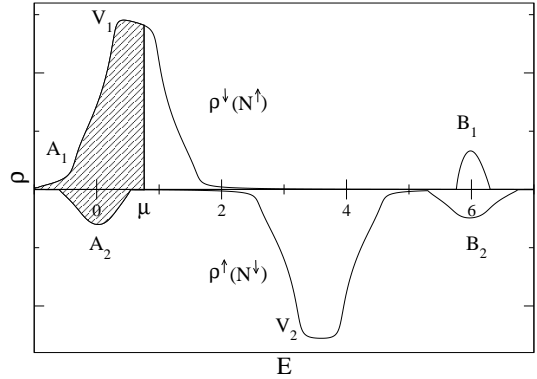}
\caption{DOS $\rho^{\up \tau CPA}$(E) and $\rho^{\down \tau CPA}$(E) using a simple cubic band 
 with $U=6$, $N^{\up}=0.1$, $N^{\down}=0.6$, $\tau=1$, $w=1$. $A_1$, $B_{1}$,  $A_2$, $B_{3}$ denote the CPA sub-bands, $V_1$ and $V_2$ the VCA bands. The shaded area marks the occupied states.\label{dosU6Ne07pict}}
\end{figure}

The transverse susceptibility $\chi_{k}^{-+}(E)$ can now be obtained as $\chi_{k}^{-+}(E)=-K_{k}^{\tau CPA}(E)$. In ferromagnetic ground states $\chi_{k}^{+-}(E)$ is also required and can be obtained similarly to $\chi_{k}^{-+}(E)$. The transitions involved in $\chi_{k}^{0 -+}(E)$ and $\chi_{k}^{0 +-}(E)$ can be understood by looking at Fig.\ref{dosU6Ne07pict}, a typical example at $\tau=1$ of a ferromagnetic equilibrium situation in the $\tau$-CPA.
$\chi_{k}^{-+}(E)$ calculates the transitions from occupied down-spin states to free up-spin states, whereas  $\chi_{k}^{+-}(E)$ is concerned with transitions from occupied up-spin states to free down-spin states. Fig.\ref{dosU6Ne07pict} shows the densities of states obtained from $G_{k}^{\up \tau CPA}(E)$ and $G_{k}^{\down \tau CPA}(E)$ calculated with the one-particle Green's functions obtained in I. In this example, $U=6.0$ and $N_e=0.7$. The equilibrium was found to be for $N^{\up}=0.1 $, $N^{\down}=0.6 $ and 
$\mu=0.76$. As is typically the case for large $U$, the chemical potential falls in the VCA band of the majority spin DOS. The sub-bands in the DOS have been labelled with $A_1$($A_2$) and $B_1$($B_2$) for the CPA sub-bands, and $V_1$($V_2$) for the VCA contribution. There is no transition between a CPA band and a VCA band. Therefore, the transitions of lowest energy happen between $V_1$ and $V_2$ in  $\chi_{k}^{-+}(E)$. At lower $U$, the CPA and VCA sub-bands combine in a complex fashion, as we have seen in I, and the CPA contributes at low energy too, though it never diverges at $E=0$.\\

The paramagnetic limit of (\ref{Kkall}) is again difficult 
to obtain, and will not be treated here. The CPA part presents the same difficulties in that case as encountered in the pure CPA. As for the VCA part, 
$m/m^{VCA}$ is varying in a complicated and self-consistent way with the species concentrations and the self-energies. It is however expected that this quantity will remain constant in the paramagnetic regime. It is therefore not unreasonable to assume that $m/m^{VCA}$ in the paramagnetic ground state retains the value it had before the transition. This value can be obtain numerically for all filling. For $N_e=0.7$, we find that $m/m^{VCA}\approx 0.8$. In the rest of this chapter we will concentrate on the application of (\ref{Kkall}) on ferromagnetic ground states. 

\subsection{\label{npm}The $n^{\pm}$ problem}
Numerical calculations of the susceptibility requires the evaluation of 
$n^{\pm}$. This value is an
equilibrium property of the system along with the chemical potential
$\mu$ for a given concentration of spins up and spins down. We recall that the weighted Green's function formalism allows for the determination of $n^{\pm}$ in the CPA, as presented in section \ref{subsub:oneGF}, through (\ref{npmOnepartCPA}). (\ref{npmOnepartCPA}) is written using the up-spins functions, but could equally be rewritten using the down-spins functions. In an equilibrium situation both formulations lead to the same solution, since the CPA leads solely to paramagnetic ground states for $N_e < 1$. In the $\tau$-CPA the system can be polarised and (\ref{npmOnepartCPA}) cannot be used.
$n^{\pm}$ cannot be deduced from the one-particle properties alone under general circumstances. In a general case $n^{\pm}$ could be determined
self-consistently, or rather $n^{-}$, since its correlation
function corresponds to the Green's function that is wanted for $\chi^{-+}$:
\begin{align}\label{npmselfcons}
n^{\pm} & = N^{-}-n^{-} \nonumber\\
n^{-}& =\langle c^{+}_{i\downarrow}c_{i\uparrow}c^{+}_{i\uparrow}
c_{i\downarrow}\rangle \nonumber \\
 & =
\int_{\mu}^{\infty}dE \left( \frac{-1}{\phantom{-}\pi}\right)
\imag \langle\langle c^{+}_{i\downarrow}
c_{i\uparrow};c^{+}_{i\uparrow}c_{i\downarrow}\rangle\rangle_{E}\nonumber\\
 &=\int_{\mu}^{\infty}dE \left(\frac{-1}{\phantom{-}\pi}\right)\imag K_{iiii}(E)
\end{align}
$K_{iiii}(E)$ depends itself on $N^{+},N^{-}$ and $n^{\pm}$
\begin{equation}\label{eqcond}
N^{\sigma}=\int_{-\infty}^{\mu}dE\left(\frac{-1}{\phantom{-}\pi}\right)\imag F^{\sigma}(E)
\end{equation}
For a fixed number of
particles, (\ref{eqcond}) is sufficient to determine $N^{+},N^{-}$
 and $\mu$, but 
(\ref{npmselfcons}, \ref{eqcond}) also form a closed set of equations with 
$n^{\pm}$ and
$\mu$ to be established self-consistently. This simple formulation hides a very long numerical process. For this
reason $n^{\pm}$ has not been obtained that way for the susceptibility calculations in the polarised regime.
For most cases of interest, such as when the system is close or
beyond the transition from paramagnetic to ferromagnetic state, the 
value of the interaction will be
fairly large. It is clear from Fig.\ref{npmcpa} that it is reasonable to put $n^{\pm}$ to
zero under those circumstances. $\Lambda$ has therefore the form given in (\ref{lambda}) and
(\ref{lambda2}), with $n^{\pm}\equiv 0$, $N^{+}\equiv n^{+}$ and
$N^{-}\equiv n^{-}$, in the numerical examples.

\subsection{\label{numsusc}Numerical results and discussion}
In this section the susceptibility is calculated using the result (\ref{Kkall}) for $\tau$-CPA at $\tau =1$ in a ferromagnetic ground state, for a simple cubic band. The most prominent features that are discussed are the spin wave spectrum and the effect of the weighted development of the residual interaction on the A-B transitions. 

\subsubsection{Spin waves}
\begin{figure}
\includegraphics[width=8.6cm]{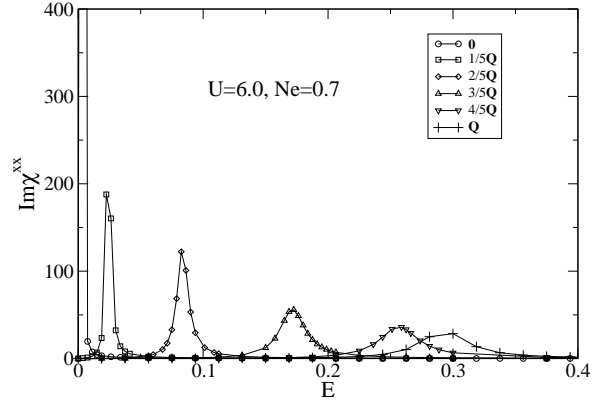}
\caption{Spin wave part of the spectrum in the ferromagnetic regime of the pure VCA for $U=6.0$, $N^{\up}=0.0$, $N^{\down}=0.7$. The imaginary part of 
$\chi^{xx}(q,E)$  is plotted for $q$ varying along the (1,1,1) direction}\label{figvcaswim6}
\end{figure}

\begin{figure}
\includegraphics[width=8.6cm,height=!]{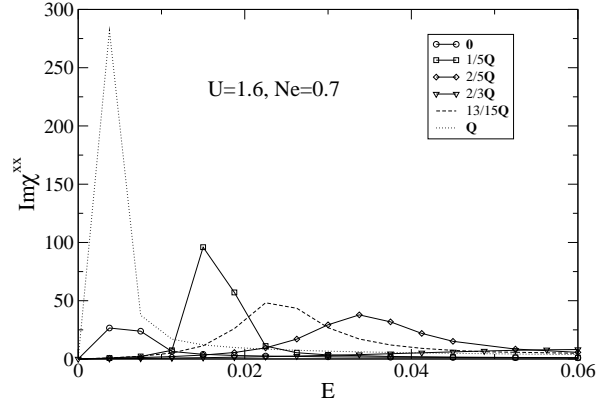}
\caption{Spin wave part of the spectrum in the ferromagnetic regime of the pure VCA for $U=1.6$, $N^{\up}=0.05$, $N^{\down}=0.65$ . The imaginary part of 
$\chi^{xx}(q,E)$  is plotted for $q$ varying along the (1,1,1) direction}\label{figvcaswim16}
\end{figure}

When the ground state of the system has become ferromagnetic, we expect to observe the formation of collective excitations starting at $q=0$ and $E=0$. Since the part of the spectrum at low energy is largely 
dominated by the VCA part, it is interesting to see first how the spin waves behave in a pure VCA case for a simple cubic band. Figs.\ref{figvcaswim6} and \ref{figvcaswim16} have been calculated using the full VCA and its equilibrium condition at $U=1.6$ and $U=6.0$ for $N_e=0.7$. 
$\chi^{xx}(q,E)$ was evaluated for $q$ along the (1,1,1) direction, from $0$ to the edge of the 
Brillouin zone at $Q$. At large $U$, Fig.\ref{figvcaswim6}, the down-spin band is full and separated by a gap from the empty up-spin band. 
The spin wave spectrum is characterised by a sharp peak 
at $q=0$. The peaks observed at finite $q$ broaden up as they enter the continuum of excitations. For the parameters of Fig.\ref{figvcaswim6}, $U=6.0$, $N_e=0.7$,  we can see that for a $q$ equal to $3/5$ Q of the lattice vector, the spin-wave is already well within the continuum.
When the ferromagnetism is not saturated, the up-spin band overlaps
with the down-spin one. Fig.\ref{figvcaswim16} is for $U=1.6$, $N^{\up}=0.05$, $N^{\down}=0.65$ and $\mu=0.28$. As the band for the down-spin is more than half-filled, its Fermi surface crosses the zone boundary, but not on the (1,1,1)-direction. 
The Fermi surface of the minority spin retains a spherically symmetric shape. The transitions allowed in calculating $\chi^{-+}$ are given by the respective positions of the two Fermi surfaces as their centres are shifted by $q$. For a simple cubic band the result is strongly dependent on the direction of $q$. Transitions at zero energy can occur when the two surfaces intersect. Fig.\ref{figvcaswim16} shows the spin-waves progressing along the energy axis. At two-thirds of $Q$, the spin-wave has completely disappeared into the continuum. At 
$q=13/15$ Q, the spin-wave has come down in energy, and though still broad, starts to emerge from the continuum. The phenomenon is of the same nature as the antiferromagnetic instability present in the mean-field phase diagram for the simple cubic crystal 
structure. 
The system in the $\tau$-CPA in this range of parameters show a tendency to become 
antiferromagnetic.
The separation of the spin-wave from the continuum is not observed, for example, along the (1,0,0) direction for the same set of parameters. \\

\begin{figure}
\includegraphics[width=8.6cm]{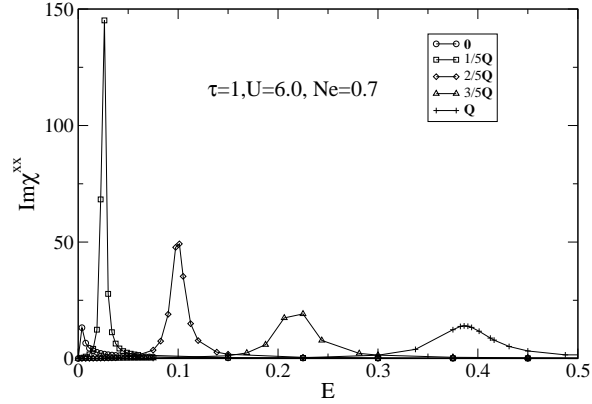}
\caption{Spin-wave part of the spectrum in the ferromagnetic regime of the $\tau$-CPA with $\tau=1$, for $U=6.0$, $N^{\up}=0.1$, $N^{\down}=0.6$, 
$\mu=0.76$. The imaginary part of 
$\chi^{xx}(q,E)$ is plotted for $q$ varying along the (1,1,1) 
direction.}\label{figswU6N07im}
\end{figure}

\begin{figure}
\includegraphics[width=8.6cm]{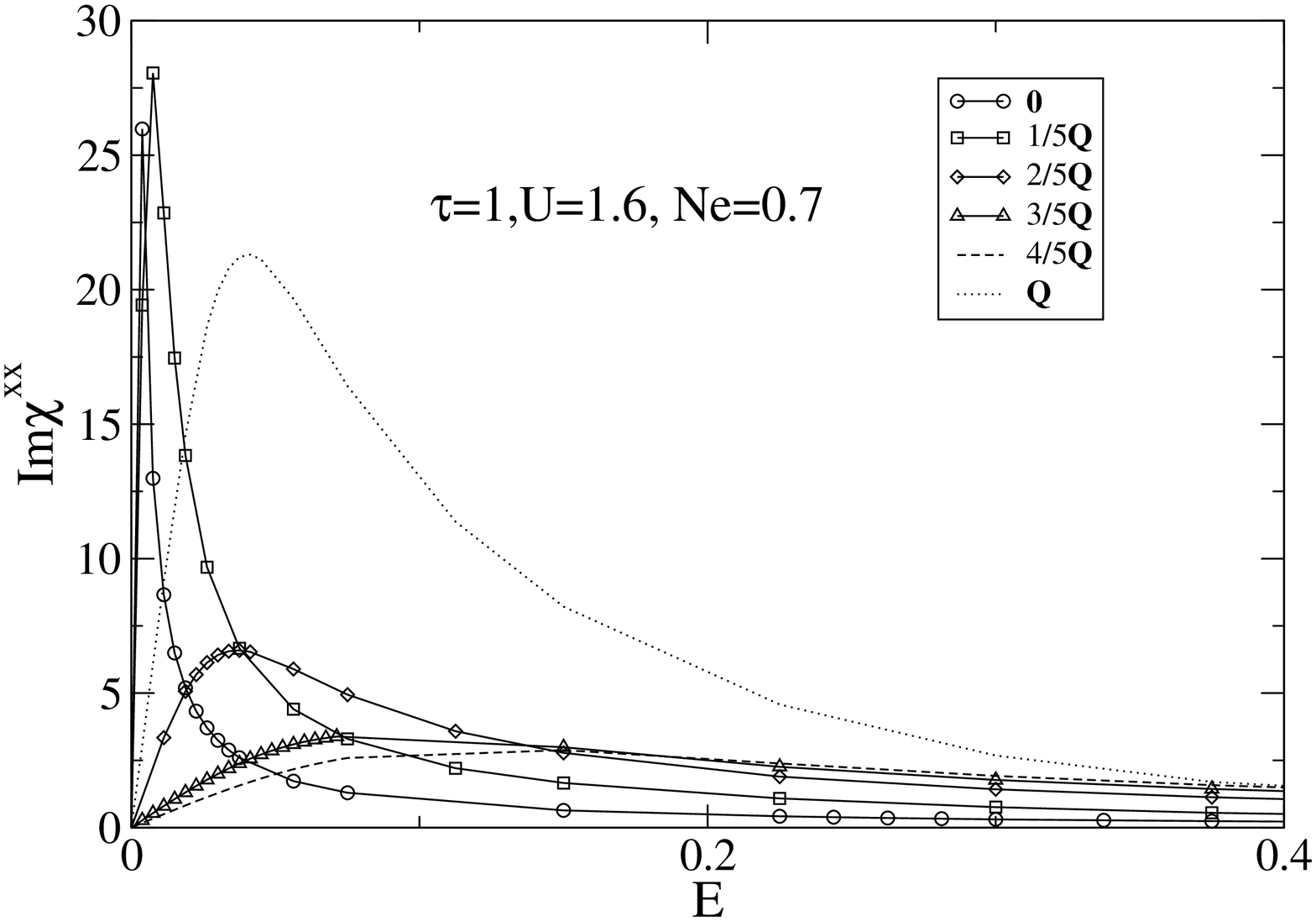}
\caption{Spin-wave part of the spectrum in the ferromagnetic regime of the $\tau$-CPA with $\tau=1$, for $U=1.6$, $N^{\up}=0.27$, $N^{\down}=0.43$, 
$\mu=0.34$. The imaginary part of 
$\chi^{xx}(q,E)$ is plotted for $q$ varying along the (1,1,1) 
direction.}\label{figswU16N07im}
\end{figure}

The system is also in a ferromagnetic ground state at $U=1.6$ and $N_e=0.7$ in the $\tau$-CPA with $\tau=1$. The equilibrium is found at $N^{\up}=0.27$ and $N^{\down}=0.43$. At $U=6.0$ and $N_e=0.7$, the system is not saturated, and $N^{\up}=0.1$, $N^{\down}=0.6$. As expected, the ferromagnetism is weaker than in the pure VCA case, but some general features remain, as can be seen in Figs.\ref{figswU6N07im} and \ref{figswU16N07im}. At $U=1.6$, the spin-waves are also seen to have decreased in energy at $q=Q$, but in that case the excitation is very broad and the peak is rather a resonance, Fig.\ref{figswU16N07im}. The $\tau$-CPA causes the general broadening of the spectrum that is well visible in Fig.\ref{figswU16N07im} and \ref{figsw0U16N07im} by forcing the VCA Green's function to match the CPA Green's function at time $\tau$.
\begin{figure}
\includegraphics[width=8.6cm]{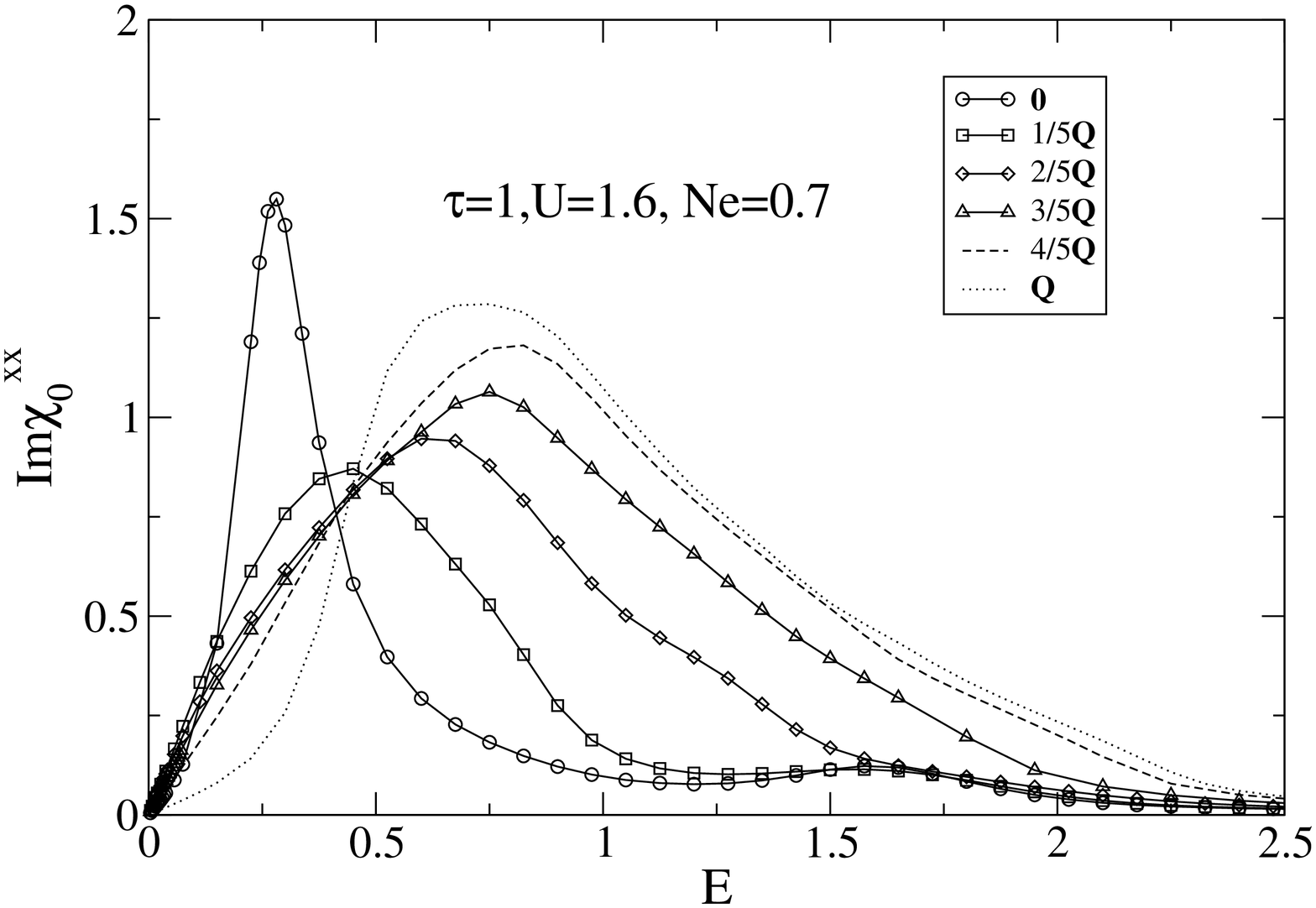}
\caption{Imaginary part of $\chi^{0 xx}(q,E)$ in the $\tau$-CPA with $\tau=1$, including the vertex correction but not the residual interaction. 
$U=1.6$, $N^{\up}=0.27$, $N^{\down}=0.43$.}\label{figsw0U16N07im}
\end{figure}
\subsubsection{The effect of the CPA part}
It has been discussed that the CPA part does not generally play a substantial part in the susceptibility at low energy. This is due to the choice of $\tau$ that gives more weight to the VCA part, 
and the resulting density of states with the chemical potential falling within the VCA part. It has also been established that the weighted correction incurred by the residual term in $U$ causes the reduction of the CPA A-B peak in the paramagnetic limit. This result is also valid in the ferromagnetic case, to the extent that this peak is completely eliminated by the weighting in $\tau$. Fig.\ref{figsw0U16N07im} show $\chi^{0}_{xx}$, before the residual interaction is taken into account. 
The CPA A-B transition band is visible at around $E=1.6$. 
\begin{figure}
\includegraphics[width=8.6cm]{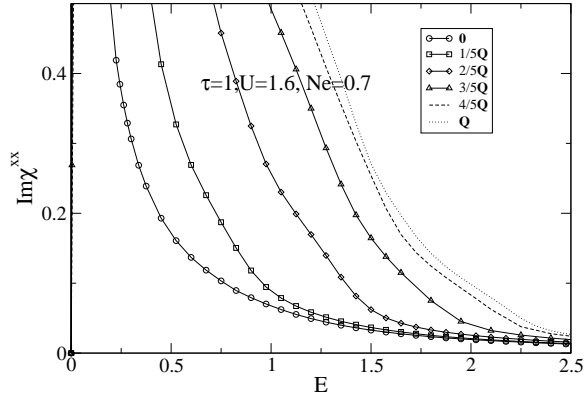}
\caption{Continuation of Fig.\ref{figswU16N07im} at larger energies illustrating the elimination of the CPA A-B peak. $\imag{\chi}^{xx}$ includes the residual interaction. $\tau=1$, $U=1.6$,$N^{\up}=0.27$, 
$N^{\down}=0.43$. }\label{ABtransred}
\end{figure}
In Fig.\ref{ABtransred}, where the residual interaction has been included, the effect is no longer visible. The broadening with $q$ at around $E=1.5$ is due to the broadening of the VCA part. Calculations for $U=6.0$ and $N_e=0.7$ show for example that the CPA A-B transitions centred at around $E=6.0$ have become almost zero for any $q$. Compared to the paramagnetic case, correlations are also generally reduced by the decreasing number of scatters common to the two species. The A-B peak is thus comparatively small even before the residual interaction is taken into account.

\section{Conclusion}
In this paper we have extended our new treatment of the magnetism in the Hubbard model using the $\tau$-CPA developed in paper I to consider the two-particle propagators and the general magnetic susceptibility of the system. This required initially the study of a more complicated system within the CPA, effectively a four-component alloy since electrons of different spins see different scattering centres, to obtain the vertex corrections. The study was then extended to obtain the one- and two-particle weighted Green's functions within the CPA approximations. Such weighted functions allow, in principle, a self-consistent calculations of $n^{\pm}$ (the concentration of doubly-occupied sites). This is only satisfactory in the paramagnetic regime, which is the only stable solution of the pure CPA. However in the $\tau$-CPA we found it necessary to use the CPA 
solutions in the ferromagnetic regime where inconsistent values of $n^{\pm}$ were found by the method. A fully self-consistent method based on the two-particle functions proved to be prohibitively complex and was not pursued, because 
for the magnetic cases of interest $U$ is relatively large, in which case $n^{\pm}$ is always very small and the difference can be neglected. \\
A full treatment of the magnetic susceptibility requires the residual interaction between pairs of electrons with opposite spins on the same site to be included. It is well known that this gives rise to spin wave excitations even 
within an RPA treatment of the Stoner (VCA) approximation. 
Within the $\tau$-CPA this requires the proper use of the weighted two-particle propagators, in part to ensure that the Goldstone mode at $q=E=0$ is preserved. Some further approximations are necessary to make the calculation tractable, which then provides a satisfactory treatment of the susceptibility. In the paramagnetic case, when treated in the appropriate limit, it shows an enhancement of transitions between sub-bands of different spins and a corresponding reduction of those between same spin bands. In the ferromagnetic case an appropriate spin wave spectrum is found.\\
Although the treatment via the $\tau$-CPA reported here is restricted to a single band in the simple cubic lattice with less than half-filling, it displays a more general theory of the magnetic state and excitations within the Hubbard model than any other previously reported. Extensions to other forms of magnetic order such as antiferromagnetism which are expected for bigger filling and more complex band structures are straightforward in principle though no doubt complex in evaluation. It should be possible to refine the approximations made within the general framework of the physical idea behind the $\tau$-CPA.

\begin{acknowledgments}
AU acknowledges the support of the
Swiss National Science Foundation, the Berrow scholarship trust of Lincoln College and the ORS award scheme.
\end{acknowledgments}

\appendix
\section{Single weight for two-particle Green's functions}\label{app:1weight}
We consider the evaluation of $\langle G^{\up n^{-}o}CG^{\down on^{-}} \rangle$. $n^{-}$ is an ``impurity'' site for the up-spin, but a ``host'' site for the down-spin propagating, so that $\langle G^{\up n^{-}o}CG^{\down on^{-}} \rangle \equiv \langle G^{\up io}CG^{\down oh} \rangle$. This in turn can be decomposed as 
\begin{equation}\label{Giooh}
\langle G^{\up io}CG^{\down oh} \rangle =\langle G^{\up io}CG^{\down} \rangle - \langle G^{\up io}CG^{\down oi} \rangle
\end{equation}
This is particular to the four-component alloy for which the two propagating particles each see a different lattice of impurities and hosts. 
The second part of (\ref{Giooh}) is already known, so that only $\langle G^{\up io}CG^{\down} \rangle$ needs to be determined. Using the definition (\ref{weightedone}), the perturbation expansion 
\begin{equation}\label{PVG_GVP} 
G =P+ PVG = P+GVP=P+PVP+PVGVP
\end{equation}
and the T-matrix expansion
\begin{equation}\label{Teq}
G=\bar{G}+\bar{G} T\bar{G}
\end{equation}
we have
\begin{equation}\label{GoiwithT}
G^{io}=\frac{1}{U}\left( P^{-1}\bar{G}+ P^{-1}\bar{G} T\bar{G}\ -1 \right) 
\end{equation}
A convenient set of equivalences are obtained from the Dyson equation 
$\bar{G}=P+P\Sigma \bar{G}$:
\begin{equation}\label{p_1G}
P^{-1}\bar{G} =\bar{G}P^{-1}= \Sigma \bar{G} +1 
\end{equation}
Putting (\ref{GoiwithT},\ref{p_1G}) into (\ref{Giooh}) gives
\begin{equation}
\langle G^{\up io}CG^{\down} \rangle = \frac{1}{U}\lbrack \Sigma^{\up} \bar{G}^{\up} C \bar{G}^{\down} +
\left( \Sigma^{\up}
\bar{G}^{\up}+1 \right) \Gamma  \bar{G}^{\down} \rbrack \label{intGioG}
\end{equation}
where Velick\'y's original expression for the vertex 
$\Gamma=\langle T^{\up}\bar{G}^{\up}C\bar{G}^{\down}T^{\down}\rangle$ appears 
in the last term of the second line of (\ref{intGioG}). The result was found to be \cite{velicky}
\begin{equation}\label{gammandef}
\Gamma_n(z_1,z_2)=\ket{n}\Lambda(z_1,z_2)
\bra{n}G^{(2)}(z_1,z_2)\ket{n}\bra{n}
\end{equation}
where $\Gamma=\sum_n \Gamma_n$, which can 
thus be substituted in (\ref{intGioG}). The diagonal element of (\ref{intGioG}) is therefore
\begin{multline}\label{GioGdiag}
\bra{m}\langle G^{\up io}CG^{\down} \rangle \ket{m} = \frac{1}{U}\lbrack \Sigma^{\up} 
\bra{m} \bar{G}^{\up} C \bar{G}^{\down}\ket{m} \\
 +\Lambda \sum_n \bra{m} ( \Sigma^{\up}
\bar{G}^{\up}+1)\ket{n}\bra{n}  \bar{G}^{\down}\ket{m}
\bra{n} \langle G^{\up}CG^{\down}\rangle \ket{n}\rbrack 
\end{multline}
The whole weighted Green's function can now be obtained by Fourier transforms. Defining
\begin{gather}
\bar{G}_{k}^{(2)}:=\sum_m e^{-ikR_m}  \bra{m}\langle G^{\up}CG^{\down} \rangle \ket{m} \label{Gk2def}\\
\bar{G}_{k}^{(2)iooo}:=\sum_m e^{-ikR_m} \bra{m}\langle G^{\up io}CG^{\down} \rangle \ket{m} \label{Gk2iooodef}\\
\bar{G}_{k}^{(2)n^{-}oon^{-}}:=\sum_m e^{-ikR_m} \bra{m}\langle G^{\up n^{-}o}CG^{\down on^{-}} \rangle \ket{m}\label{Gk2nmoomndef}
\end{gather}
 The value of $\bar{G}_{k}^{(2)}$ is already known, (\ref{Kkz1z2velick}). The result for $\bar{G}_{k}^{(2)iooo} $ is 
\begin{equation}\label{Gk2iooores}
\bar{G}_{k}^{(2)iooo}=\frac{1}{U}\left[ \Sigma^{\up}a_k +
\Lambda \left( \Sigma^{\up}A_k + F^{\down} \right)\bar{G}_{k}^{(2)} \right]
\end{equation}
The Fourier transforms $a_k$ and $A_k$ were defined in (\ref{dekak}) and (\ref{defAk}).
Gathering the results (\ref{Giooh}), (\ref{G2npmoonpm}) and (\ref{Gk2iooores}), we have finally
\begin{equation}\label{Gknm00nm}
\bar{G}_{k}^{(2)n^{-}oon^{-}}=\bar{G}_{k}^{(2)}\left(\frac{\Sigma^{\up}+\Lambda F^{\down}}{U}-\frac{\Sigma^{\up}\Sigma^{\down}+\Lambda D }{U^2}  \right)
\end{equation} 
Similar calculations can be carried out to obtain $\langle G^{\up n^{+}o} CG^{\down on^{+}} \rangle$ from $\langle G^{\up ho} CG^{\down oi} \rangle$, and $\langle G^{\up n^{\varnothing}o} CG^{\down on^{\varnothing}} \rangle$ from $\langle G^{\up ho} CG^{\down oh} \rangle$. We find
\begin{gather}
\bar{G}_{k}^{(2)n^{+}oon^{+}} =\bar{G}_{k}^{(2)}\left(\frac{\Sigma^{\down}+\Lambda F^{\up}}{U}-\frac{\Sigma^{\up}\Sigma^{\down}+\Lambda D }{U^2}  \right) \\
\begin{split}
&\bar{G}_{k}^{(2)n^{\varnothing}oon^{\varnothing}} =\bar{G}_{k}^{(2)}\\
&\times \left(1-\frac{\Sigma^{\down}+\Lambda F^{\up}}{U}-\frac{\Sigma^{\up} + 
\Lambda F^{\down}}{U} +
\frac{\Sigma^{\up}\Sigma^{\down}+\Lambda D }{U^2}  \right) 
\end{split}\label{G2n0oon0}
\end{gather}
It is more difficult to calculate $\langle G^{\up on^{-}}CG^{\down n^{-}o} \rangle $, with the weights swapped inside, as $\Gamma_1=\langle 
T^{\up}  \bar{G}^{\up} P^{-1}C \bar{G}^{\down} T^{\down}\rangle  $ appears in the evaluation of $\langle G^{\up oi}CG^{\down} \rangle$ instead of $\Gamma$. $\Gamma_1$ will be evaluated in the context of the double weights in the next section. With the knowledge of $\Gamma$ it can be shown that for $C$ diagonal, $\langle G^{\up oi}CG^{\down} \rangle\equiv\langle G^{\up io}CG^{\down} \rangle$. Elliott and Schwabe found that in this case, $\langle G^{\up oi}CG^{\down io} \rangle$ is also equivalent to  $\langle G^{\up io}CG^{\down oi} \rangle$, so that $\langle G^{\up on^{-}}CG^{\down n^{-}o} \rangle \equiv \langle G^{\up n^{-}o}CG^{\down on^{-}} \rangle $. This result is general for the other single weighting too. Using the expression for the vertex correction (\ref{lambda}), the weights in 
 (\ref{G2npmoonpm}) and (\ref{Gknm00nm}-\ref{G2n0oon0}) can be rearranged and give the results detailed in (\ref{singleweights}).

\section{Double weight for two-particle Green's functions}\label{app:2weight}
The case $\langle 
G^{\up n^{-} n^{-}}CG^{\down n^{-} n^{-} } \rangle $ is equivalent to 
$\langle 
G^{\up ii}CG^{\down hh } \rangle $. The Green's function sought after can be expressed as
\begin{align}\label{develGiiGhh}
\langle G^{\up ii}CG^{\down hh } \rangle &=\langle G^{\up ii}CG^{\down} \rangle-\langle G^{\up ii}CG^{\down oi } \rangle \nonumber \\
 &-\langle G^{\up ii}CG^{\down io } \rangle+\langle G^{\up ii}CG^{\down ii } \rangle
\end{align}
The last weighted Green's function on the right-hand side is already known, but the three others must be separately determined. Using the perturbation expansion (\ref{PVG_GVP}) and the T-matrix expansion (\ref{Teq}), 
\begin{equation}\label{Giinoav}
 G^{\up ii} = \frac{1}{U^2}\left(
P^{-1}\bar{G} P^{-1}+P^{-1} \bar{G} T \bar{G}P^{-1}-P^{-1}-V \right)
\end{equation}
The development (\ref{Giinoav}) can be substituted into the second term in the 
right-hand side of (\ref{develGiiGhh}).  After having eliminating the contribution containing $\langle T \rangle$, and applying (\ref{p_1G}), (\ref{Giinoav}) becomes
\begin{gather}
\langle G^{\up ii}CG^{\down oi } \rangle = \mathcal{M}_1+\mathcal{K}_1 \label{Giioidev}\\
\mathcal{M}_1= -\frac{1}{U^3}\langle V^{\up}\rangle C \Sigma^{\down}\bar{G}^{\down}-\frac{1}{U^3}
\langle V^{\up}C\bar{G}^{\down} T^{\down} \rangle \bar{G}^{\down} P^{-1}\\
\mathcal{K}_1=\frac{1}{U^3}\left( \Sigma^{\up}\Sigma^{\down} P^{-1} \bar{G}^{\up}C \bar{G}^{\down}+P^{-1} \bar{G}^{\up}\Gamma_1 \bar{G}^{\down} P^{-1} \right) \label{K1}\\
\Gamma_1:=\langle 
T^{\up}  \bar{G}^{\up} P^{-1}C \bar{G}^{\down} T^{\down}\rangle \label{gamma_1}
\end{gather}
where $V^{\sigma}$ is the interaction matrix for the spin $\sigma$. 
The terms involving the disorder matrix $V^{\up}$ are gathered in $ \mathcal{M}_1 $. As discussed by 
Elliott and Schwabe \cite{schwabe1}, for single-site disorder the matrices $V^{\up}$ and $V^{\down}$ are diagonal, and all contributions from $\mathcal{M}_1(i,j)$ vanish for $i \neq j$ when (\ref{Giioidev}) is projecting onto the real space. The total contribution from $\mathcal{M}_1$ is an order of magnitude smaller in $N$ than $\mathcal{K}_1$ and can thus be neglected. Unlike the singly-weighted case (\ref{intGioG}), the vertex correction does not appear as such in the second term of $\mathcal{K}_1$ (\ref{K1}). In order to obtain 
from (\ref{K1}) a self-consistent equation, we develop the average as was done by Velick\'y for the vertex correction, using the effective wave approximation results and the definitions of 
$Q$ and $\widetilde{Q}$
(Eqs.(33) to (36) in Velick\'y \cite{velicky}) and get
\begin{equation}\label{defgamma1}
\Gamma_1 =\sum_{nm}\langle Q^{\up}_n \bar{G}^{\up}  P^{-1} C \bar{G}^{\down}
\widetilde{Q}_{m}^{\down}
\rangle := \sum_{n}\Gamma_{1\,n}
\end{equation}
Following through the procedure detailed in Ref.\onlinecite{velicky}, we find
\begin{align}\label{G1ndetail}
\Gamma_{1\,n} &= \langle T_{n}^{\up} \bar{G}^{\up}P^{-1}C \bar{G}^{\down}T_{n}
^{\down} \rangle + \langle T_{n}^{\up} \bar{G}^{\up} \Gamma_1 \bar{G}^{\down}T_{n}
^{\down} \rangle \nonumber \\
 &-\langle T_{n}^{\up} \bar{G}^{\up} \Gamma_{1\,n} \bar{G}^{\down}T_{n}
^{\down} \rangle
\end{align}
From (\ref{K1}), $\bar{G}^{\up} \Gamma_{1\,n} \bar{G}^{\down}$ can be extracted as 
\begin{equation}\label{GgammaG}
\bar{G}^{\up} \Gamma_{1\,n} \bar{G}^{\down}=U^3 P \mathcal{K}_1 P -\Sigma^{\up}
\Sigma^{\down}\bar{G}^{\up} C \bar{G}^{\down} P
\end{equation}
which leads to
\begin{align}
\Gamma_{1\, n} &=\langle T_{n}^{\up}\bar{G}^{\up}P^{-1}C \bar{G}^{\down}T_{n}^{\down} \rangle +
 U^3 \langle T_{n}^{\up} P\mathcal{K}_1 P T_{n}^{\down} \rangle \nonumber \\
 &-  \langle T_{n}^{\up}\bar{G}^{\up}\Gamma_{1\, n}\bar{G}^{\down}T_{n}^{\down} \rangle-\Sigma^{\up}\Sigma^{\down}  \langle T_{n}^{\up}\bar{G}^{\up}C \bar{G}^{\down} P T_{n}^{\down} \rangle\label{K1calc}
\end{align}
Under this form, introducing the site projection $T_m=\ket{m}t_m\bra{m}$ allows us to extract $\Lambda$ from $\Gamma_{1\, n}$
\begin{gather}
\Gamma_{1\, n} =\ket{n}\Lambda \bra{n}\lbrace \bra{n}\bar{G}^{\up}P^{-1}C \bar{G}^{\down} 
\ket{n} \nonumber \\
 + U^3   \bra{n}P \mathcal{K}_1 P \ket{n} -\Sigma^{\up}
\Sigma^{\down} \bra{n}  \bar{G}^{\up} C \bar{G}^{\down} P \ket{n}
\rbrace \label{gamma1site}
\end{gather}
By replacing (\ref{gamma1site}) into the original equation for $\mathcal{K}_1
$, (\ref{K1}), it is possible to solve the equation for $\langle P \mathcal{K}_1 P \rangle $. The following Fourier transforms are defined:
\begin{gather}
\tilde{\mathcal{K}}_{1\,k}=U^{3}\sum_m e^{-ikR_m}\bra{m}P \mathcal{K}_1 P \ket{m} \\
\tilde{a}_k=\sum_m e^{-ikR_m}\bra{m} \bar{G}^{\up} C \bar{G}^{\down} P\ket{m} 
\end{gather}
Along with the definitions (\ref{dekak},\ref{defAk}) of $a_k$ and $A_k$, the solution reads
\begin{equation}
\tilde{\mathcal{K}}_{1\,k}=\frac{\Sigma^{\up}
\Sigma^{\down} \tilde{a}_k + \Lambda A_k\left(\Sigma^{\up}a_k +F^{\down}-
 \Sigma^{\up}
\Sigma^{\down}  \tilde{a}_k \right) }{1-\Lambda A_k}
\end{equation}
The complete solution for $\mathcal{K}_1(k)=\sum_m e^{-ikR_m}\bra{m} \mathcal{K}_1\ket{m}$ can finally be found, after some algebraic manipulation
\begin{equation}\label{mathcalK1}
\mathcal{K}_1(k)=G_{k}^{2}\left(\frac{ \Sigma^{\up}\Sigma^{\down}+\Lambda D}{U^2}\right) \left(  \frac{\Sigma^{\up}}{U} +\frac{F^{\down}}{a_k U}  \right)
\end{equation}
As Elliott and Schwabe commented \cite{schwabe1}, it is consistent with the assumption that $\mathcal{M}_1$ is negligible to ignore the k-independent term. In the above  (\ref{mathcalK1}), if the  k-independent term $F^{\down}\left( \Lambda D +  \Sigma^{\up}\Sigma^{\down}\right)$ is neglected, it reduces to the more convenient form
\begin{equation}\label{mathcalK1b}
\mathcal{K}_1(k)=G_{k}^{2}\left( \frac{\Sigma^{\up}\Sigma^{\down}+\Lambda D}{U^2}\right) \left(  \frac{\Sigma^{\up}+\Lambda F^{\down}}{U} \right)
\end{equation} 
Since $\mathcal{M}_1$ is neglected, (\ref{mathcalK1b}) is also the value for (\ref{Giioidev}), $G^{2 \,iioi }$. \\
A very similar calculation yields the same result for $ G^{2 \,iiio }$, $G^{2 \,iioi } \equiv G^{2 \,iiio }$. $G^{2 \,iioo }$ can be obtained along the same lines too, giving
\begin{equation}\label{G2iioo}
G_{k}^{2 \,iioo }=G_{k}^{2}\left( \frac{\Sigma^{\up} +\Lambda F^{\down}}{U}\right)^2 
\end{equation} 
Combining (\ref{G2iiii}), (\ref{mathcalK1b}) and (\ref{G2iioo}) and into (\ref{develGiiGhh}) gives
\begin{equation}
G_{k}^{(2) \, n^{-}n^{-}n^{-}n^{-} }=\left( \frac{\Sigma^{\up} +\Lambda F^{\down}}{U}-\frac{\Sigma^{\up}\Sigma^{\down}+\Lambda D}{U^2}  \right)^2 
\end{equation}
So we conclude that the weight of the doubly-weighted two-particle Green's functions is the multiplication of the simple weights, provided that $\mathcal{M}_1$ can be ignored and (\ref{mathcalK1b}) is correct.

\bibliographystyle{apsrev2} 
\bibliography{biblio2}

\end{document}